\documentclass[10pt]{article}

\usepackage{amsmath}
\usepackage{amssymb}

\usepackage{graphicx}

\usepackage{cite}

\usepackage{color} 

\usepackage[labelfont=bf,labelsep=period,justification=raggedright]{caption}

\makeatletter
\renewcommand{\@biblabel}[1]{\quad#1.}
\makeatother

\date{}

\begin{document}

\begin{flushleft}
{\Large
\textbf{From principal component to direct coupling analysis of coevolution in 
proteins: Low-eigenvalue modes are needed for structure prediction}
}
\\
$\mathrm{Simona\ Cocco}^{1}$, 
$\mathrm{Remi\ Monasson}^{2}$, 
$\mathrm{Martin\ Weigt}^{3,4,\ast}$
\\
\bf{1} Laboratoire de Physique Statistique de l'Ecole Normale
  Sup\'erieure - UMR 8550, associ\'e au CNRS et \`a l'Universit\'e
  Pierre et Marie Curie, 24 rue Lhomond, 75005 Paris,France
\\
\bf{2} Laboratoire de Physique Th\'eorique de l'Ecole Normale
  Sup\'erieure - UMR 8549, associ\'e au CNRS et \`a l'Universit\'e
  Pierre et Marie Curie, 24 rue Lhomond, 75005 Paris, France
\\
\bf{3} Universit\'e Pierre et Marie Curie, UMR 7238 -
  Laboratoire de G\'enomique des Microorganismes, 15 rue de l'Ecole de
  M\'edecine, 75006 Paris, France
\\
\bf{4} Human Genetics Foundation, Via Nizza 52, 10126 Torino,
  Italy
\\
$\ast$ E-mail: martin.weigt@upmc.fr
\end{flushleft}

\section*{Abstract}
Various approaches have explored the covariation of residues in
multiple-sequence alignments of homologous proteins to extract
functional and structural information. Among those are principal
component analysis (PCA), which identifies the most correlated groups
of residues, and direct coupling analysis (DCA), a global inference
method based on the maximum entropy principle, which aims at
predicting residue-residue contacts. In this paper, inspired by the
statistical physics of disordered systems, we introduce the
Hopfield-Potts model to naturally interpolate between these two
approaches. The Hopfield-Potts model allows us to identify relevant
'patterns' of residues from the knowledge of the eigenmodes and
eigenvalues of the residue-residue correlation matrix. 
We show how the computation of such statistical patterns makes it
possible to accurately predict residue-residue contacts with a much
smaller number of parameters than DCA. This dimensional reduction
allows us to avoid overfitting and to extract contact information from
multiple-sequence alignments of reduced size. In addition, we show that
low-eigenvalue correlation modes, discarded by PCA, are important to
recover structural information: the corresponding patterns are highly
localized, that is, they are concentrated in few sites, which we
find to be in close contact in the three-dimensional protein fold.
\section*{Author Summary}
Extracting functional and structural information about protein
families from the covariation of residues in multiple sequence
alignments is an important challenge in computational biology. Here we
propose a statistical-physics inspired framework to analyze those
covariations, which naturally unifies existing methods in the
literature. Our approach allows us to identify statistically relevant
'patterns' of residues, specific to a protein family. We show that
many patterns correspond to a small number of sites on the protein
sequence, in close contact on the 3D fold. Hence, those patterns allow
us to make accurate predictions about the contact map from sequence
data only. Further more, we show that the dimensional reduction, which
is achieved by considering only the statistically most significant
patterns, avoids overfitting in small sequence alignments, and improves
our capacity of extracting residue contacts in this case.

\section*{Introduction}
\label{sec:intro}

Thanks to the constant progresses in DNA sequencing techniques, by now
more than 4,400 full genomes are sequenced \cite{gold}, resulting in
more than $3.6\ 10^7$ known protein sequences \cite{uniprot}, which
are classified into more than 14,000 protein domain families
\cite{pfam}, many of them containing in the range of $10^3-10^5$
homologous ({\it i.e.} evolutionarily related) amino-acid sequences. These
huge numbers are contrasted by only about 92,000 experimentally
resolved X-ray or NMR structures \cite{pdb}, many of them describing
the same proteins. It is therefore tempting to use sequence data alone
to extract information about the functional and the structural
constraints acting on the evolution of those proteins. Analysis of
single-residue conservation offers a first hint about those
constraints: Highly conserved positions (easily detectable in multiple
sequence alignments corresponding to one protein family) identify
residues whose mutations are likely to disrupt the protein function,
{\it e.g.} by the loss of its enzymatic properties. However, not all
constraints result in strong single-site conservation. As is
well-known, compensatory mutations can happen and preserve the
integrity of a protein even if single site mutations have deleterious
effects \cite{gobel94,lockless99}. A natural idea is therefore to
analyze covariations between residues, that is, whether their
variations across sequences are correlated or not \cite{valentia2013}.
In this context, one introduces a matrix $\Gamma_{ij}(a,b)$ of
residue-residue correlations expressing how much the presence of
amino-acid '$a$' in position '$i$' on the protein is correlated across
the sequence data with the presence of another amino-acid '$b$' in
another position '$j$'. Extracting information from this matrix has
been the subject of numerous studies over the past two decades, see
{\it e.g.}
\cite{gobel94,casari95,ortiz97,panzos97,ortiz99,lockless99,lapedes02,fodor04,socolich05,russ05,dunn08,weigt09,burger10,morcos11,balakrishnan11,jones12}
and \cite{valentia2013} for a recent up-to-date review of the field.
In difference to these correlation-based approaches, Yeang {\it et
  al.}~\cite{haussler07}, proposed a simple evolutionary model which
measures coevolution in terms of deviation from independent-site
evolution. However, a full dynamical model for residue coevolution is
still outstanding.

The direct use of correlations for discovering structural constraints
such as residue-residue contacts in a protein fold has, unfortunately,
remained of limited accuracy
\cite{gobel94,ortiz97,ortiz99,lockless99,fodor04,dunn08}.  More
sophisticated approaches to exploit the information included in
$\Gamma$ are based on a {\it Maximum Entropy} (MaxEnt)
\cite{jaynes57a,jaynes57b} modeling. The underlying idea is to look
for the simplest statistical model of protein sequences capable of
reproducing empirically observed correlations. MaxEnt has been used to
analyze many types of biological data, ranging from multi-electrode
recording of neural activities \cite{schneidman06,cocco09}, gene
concentrations in genetic networks \cite{Lezon2006}, bird flocking
\cite{Bialek2012} etc. MaxEnt to model covariation in protein
sequences was first proposed in a purely theoretical setting by
Lapedes {\it et al.}~\cite{lapedes99}, and applied to protein sequences
in an unpublished preprint by 
Lapedes {\it et al.}~\cite{lapedes02}. It was used -- even if not
explicitly stated -- by Ranganathan and coworkers to generate random
protein sequences through Monte Carlo simulations, as a part of an
approach called Statistical Coupling Analysis (SCA) \cite{russ05}. 
Remarkably, many of those artificial proteins folded into a
native-like state, demonstrating that
MaxEnt modeling was able to statistically capture essential features
of the protein family. Recently, one of us proposed, in a series of
collaborations, two analytical approaches based on mean-field type
approximations of statistical physics, called {\it Direct Coupling
  Analysis} (DCA), to efficiently compute and exploit this MaxEnt
distribution (\cite{weigt09} uses message passing, \cite{morcos11}
a computationally more efficient naive mean-field approximation), 
related approaches developed partially in parallel are
\cite{burger10,balakrishnan11,jones12}. Informally speaking, DCA
allows for disentangling direct contributions to correlations
(resulting from native contacts) from indirect contributions 
(mediated through chains of native contacts). Hence, DCA offers a 
much more accurate image of the
contact map than $\Gamma$ itself. The full potential of maximum-entropy
modeling for accurate structural prediction was first recognized in 
\cite{schug09} (quaternary structure prediction) and in
\cite{marks11} (tertiary structure prediction), and further applied by
\cite{sadowski11,sulkowska12,nugent12,hopf12,dago12,marks12,taylor13}.
It became obvious that the extracted information is sufficient to
predict folds of relatively long proteins and transmembrane domains.
In \cite{dago12} it was used to rationally design mutagenesis
experiments to repair a non-functional hybrid protein, and thus to
confirm the predicted structure.

Despite its success, MaxEnt modeling raises several
concerns. The number of 'direct coupling' parameters necessary to
define the MaxEnt model over the set of protein sequences, is of the
order of $L^2(q-1)^2$. Here, $L$ is the protein length, and $q=21$ is
the number of amino acids (including the gap). So, for realistic
protein lengths of $L=50-500$, we end up with $10^6-10^8$ parameters,
which have to be inferred from alignments of $10^3-10^5$
proteins. Overfitting the sequence data is therefore a major risk.

Another mathematically simpler way to extract information from
the correlation matrix $\Gamma$ is Principal Component Analysis (PCA)
\cite{Pearson1901}. PCA looks for the eigenmodes of $\Gamma$
associated to the largest eigenvalues. These modes are the ones
contributing most to the covariation in the protein family. Combined
with clustering approaches, PCA was applied to identify functional
residues in \cite{casari95}.  More recently PCA was applied to the SCA
correlation matrix, a variant of the matrix $\Gamma$ expressing
correlations between sites only (and not explicitly the amino-acids
they carry) and allowed for identifying groups of correlated
(coevolving) residues -- termed sectors -- each controlling a specific
function \cite{halabi09}.  A fundamental
issue with PCA is the determination of the number of relevant
eigenmodes. This is usually done by comparing the spectrum of $\Gamma$
with a null model, the Marcenko-Pastur (MP) distribution, describing
the spectral properties of the sample covariance matrix of a set of
independent variables \cite{bai2009}. Eigenvalues larger than the top
edge of the MP distribution cannot be explained from sampling noise
and are selected, while lower eigenvalues -- inside the bulk of the MP
spectrum, or even lower -- are rejected.

In this article we show that there exists a deep connection between
DCA and PCA. To do so we consider the Hopfield-Potts model, an
extension of the Hopfield model introduced three decades ago in
computational neuroscience \cite{hopfield82}, to the case of variables
taking $q>2$ values. The Hopfield-Potts model is based on the concept
of patterns, that is, of special directions in sequence
space. These patterns show some similarities with sequence motifs or
position-specific scoring matrices, but instead of encoding
independent-site amino-acid preferences, they include statistical
couplings between sequence positions. Some of these patterns are
'attractive', defining 'ideal' sequence motifs which real sequences in
the protein family try to mimic.  In distinction to the
original Hopfield model \cite{hopfield82}, we also find 'repulsive'
patterns, which define regions in the sequence space deprived of real
sequences. The statistical mechanics of the inverse Hopfield model,
studied in \cite{CoMoSe} for the $q=2$ case and extended here to the
generic $q>2$ Potts case, shows that it naturally interpolates between
PCA and DCA, and allows us to study the statistical issues raised by
those approaches exposed above. We show that, in contradistinction
with PCA, low eigenvalues and eigenmodes are important to recover
structural information about the proteins, and should not be
discarded. In addition, we propose a maximum likelihood criterion for
pattern selection, not based on the comparison with the MP
spectrum. We study the nature of the statistically most significant 
eigenmodes, and show that they exhibit remarkable
features in term of localization: most repulsive patterns are strongly
localized on a few sites, generally found to be in
close contact on the three-dimensional structure of the proteins. As
for DCA, we show that the dimensionality of the MaxEnt model can be
very efficiently reduced with essentially no loss of predictive power
for the contact map in the case of large multiple-sequence alignments,
and with an improved accuracy in the case of small alignments containing
too few sequences for standard mean-field DCA to work.
These conclusions are established both from
theoretical arguments, and from the direct application of the
Hopfield-Potts model to a number of sample protein families.

\subsection*{A short reminder of covariation analysis}
\label{sec:mf}

Data are given in form of a {\it multiple sequence alignment} (MSA), in 
which each row contains the amino-acid sequence of one protein, and each
column one residue position in these proteins, which is aligned based 
on amino-acid similarity. We denote the MSA by $A = \{ a_i^m |
i=1,...,L,\ m=1,...,M\}$ with index $i$ running over the $L$ columns of 
the alignment (residue positions / sites) , and $m$ over the $M$
sequences, which constitute the rows of the MSA. The amino-acids $a_i^m$ 
are assumed to be represented by natural numbers $1,...,q$
with $q=21$, where we include the 20 standard amino acids and the 
alignment gap '-'.

In our approach, we do not use the data directly, but we summarize
them by the amino-acid occupancies in single columns and pairs of
columns of the MSA (cf. Methods for data preprocessing),
\begin{eqnarray}
f_i(a) &=& \frac 1{M} \sum_{m=1}^M \delta_{a,a_i^m} \\
f_{ij}(a,b) &=& \frac 1{M} 
\sum_{m=1}^M \delta_{a,a_i^m}\delta_{b,a_j^m} \ ,
\label{eq:fij}
\end{eqnarray}
with $i,j=1,...,L$ and $a,b=1,...,q$. The Kronecker symbol
$\delta_{a,b}$ equals one for $a=b$, and zero else. Since frequencies
sum up to one, we can discard one amino-acid value ({\it e.g.}~$a=q$)
for each position without losing any information about the sequence
statistics. We define the empirical covariance matrix through
\begin{equation}
C_{ij}(a,b) = f_{ij}(a,b) - f_i(a) f_j(b) \ ,
\label{eq:cov}
\end{equation}
with the position index $i$ running from $1$ to $L$, and the
amino-acid index from $1$ to $q-1$. The covariance matrix $C$ can
therefore be interpreted as a square matrix with $(q-1)L$ rows 
and columns. We will adopt this interpretation throughout the paper,
since the methods proposed become easier in terms of the linear 
algebra of this matrix.

\subsubsection*{Maximum entropy modeling and direct couplings}

Non-zero covariance between two sites does not necessarily imply 
the sites to directly
interact for functional or structural purposes \cite{fodor04}. The
reason is the following \cite{weigt09}: When $i$ interacts with $j$,
and $j$ interacts with $k$, also $i$ and $k$ will show correlations
even it they do not interact. It is thus important to distinguish
between {\it direct} and {\it indirect} correlations, and to infer
{\it networks of direct couplings}, which generate the empirically
observed covariances. This can be done by constructing a
(protein-family specific) statistical model $P(a_1,...,a_L)$, which
describes the probability of observing a particular amino-acid
sequence $a_1,...,a_L$. Due to the limited amount of available data,
we require this model to reproduce empirical frequency counts for
single MSA columns and column pairs,
\begin{eqnarray}\label{eq:coherent}
f_i(a_i) &=& \sum_{\{a_k|k\neq i\}} P(a_1,...,a_L) \\
f_{ij}(a_i,a_j) &=& \sum_{\{a_k|k\neq i,j\}} P(a_1,...,a_L)\ , \label{eq:f2}
\end{eqnarray}
{\it i.e.}~marginal distributions of $P(a_1,...,a_L)$ are required to
coincide with the empirical counts up to the level of position
pairs. Beyond this coherence, we aim at the {\it least constrained}
statistical description. The {\it maximum-entropy principle}
\cite{jaynes57a,jaynes57b} stipulates that $P$ is found by maximizing
the entropy
\begin{equation}
H[P] = -\sum_{a_1,...,a_L} P(a_1,...,a_L) \log P(a_1,...,a_L) \ ,
\label{eq:maxent}
\end{equation}
subject to the constraints Eqs.~(\ref{eq:coherent}) and (\ref{eq:f2}).
We readily find the analytical form
\begin{equation}
P(a_1,...,a_L) = \frac 1{{\cal Z}(\{e_{ij}(a,b),h_i(a)\})} \exp\left\{
\frac 12 \sum_{i,j}e_{ij}(a_i,a_j) + \sum_i h_i(a_i)
\right\}\ ,
\label{eq:potts}
\end{equation}
where $\cal Z$ is a normalization constant. The MaxEnt model thus
takes the form of a (generalized) $q$-states Potts model, a celebrated
model in statistical physics \cite{wu}, or a Markov random field
in a more mathematical language. The parameters $e_{ij}(a,b)$
are the direct couplings between MSA columns, and the $h_i(a)$
represent the local fields (position-weight matrices) acting on single 
sites.  Their values have to be determined such that 
Eqs.~(\ref{eq:coherent}) and (\ref{eq:f2}) are satisfied.
Note that, without the coupling terms $e_{ij}(a,b)$, the
model would reduce to a standard position-specific scoring
matrix. It would describe independent sites, and thus it would
be intrinsically unable to capture residue covariation.

From a computational point of view, however, it is not possible to
solve Eqs.~(\ref{eq:coherent}) and (\ref{eq:f2}) exactly. The reason
is that the calculations of ${\cal Z}$ and of the marginals require
summations over all $q^L$ possible amino-acid sequences of length 
$L$. With $q=21$ and
typical protein lengths of $L=50-500$, the numbers of configurations
are enormous, of the order of $10^{65}-10^{650}$. The way out is an
approximate determination of the model parameters. The computationally
most efficient way found so far is an approximation, called mean field
in statistical physics, leading to the approach known as {\em direct
  coupling analysis} \cite{morcos11}. Within this mean-field
approximation, the values for the direct couplings are simply equal to
\begin{equation}
e_{ij}(a,b) = (C^{-1})_{ij}(a,b)
\ \ \ \ \ \ \forall i,j\ \forall a,b=1,\ldots, q-1,
\label{eq:nmf}
\end{equation}
and $e_{ij}(a,q)=e_{ij}(q,a)=0$ for all $a=1,\ldots, q$. Note that the
couplings can be approximated with this formula in a time of the order
of $L^3(q-1)^3$, instead of the exponential time complexity, $q^L$,
of the exact calculation. On a single desktop PC, this can be
achieved in a few seconds to minutes, depending on the length $L$ of
the protein sequences.

The problem can be formulated equivalently in terms of
maximum-likelihood (ML) inference. Assuming $P(a_1,..,a_L)$ to be a
pairwise model of the form of Eq.~(\ref{eq:potts}), we aim at
maximizing the log-likelihood
\begin{equation}
{\cal L}\left[ \{e_{ij}(a,b),h_i(a)\} | A \right] =
\frac 1M \sum_{m=1}^M \log P(a_1^m,...,a_L^m)
\label{eq:ll_intro}
\end{equation}
of the model parameters $\{e_{ij}(a,b),h_i(a)\}$ given the MSA $A$.
This maximization implies that Eqs.~(\ref{eq:coherent}) and
(\ref{eq:f2}) hold. In the rest of the paper, we will adopt the
point of view of ML inference, cf.~the details given in Methods.
Note that, without restrictions on the couplings $e_{ij}(a,b)$
ML and MaxEnt inference are equivalent, but under the specific
form for $e_{ij}(a,b)$ assumed in the Hopfield-Potts model, this
equivalence will break down. More precisely, the ML model will fit  
Eqs.~(\ref{eq:coherent}) and (\ref{eq:f2}) only approximately

Once the direct couplings $e_{ij}(a,b)$ have been calculated, they can
be used to make predictions about the contacts between residues, details 
can be found in the Methods
Section. In \cite{morcos11}, it was shown that the predictions for the
residue-residue contacts in proteins are very accurate. In other
words, DCA allows to find a very good estimate of a partial contact
map from sequence data only. Subsequent works have shown that this
contact map can be completed by embedding it into three dimensions
\cite{marks11,sulkowska12}.
 
\subsubsection*{Pearson correlation matrix and principal component analysis}

Another way to extract information about groups of correlated residues
is the following.  From the covariance matrix $C$ given in
Eq.~(\ref{eq:cov}), we construct the Pearson correlation matrix
$\Gamma$ through the relationship
\begin{equation}
\Gamma_{ij}(a,b) = \sum_{c,d=1}^{q-1} (D_i)^{-1}(a,c)\;
C_{ij}(c,d)\;  (D_j)^{-1}(d,b)\ ,
\label{eq:gamma}
\end{equation}
where the matrices $D_i$ are the square roots of the single-site
correlation matrices, {\it i.e.}
\begin{equation}
C_{ii}(a,b) = \sum_{c=1}^{q-1} D_i(a,c) D_i(c,b)\ .
\label{eq:Diab}
\end{equation}
This particular form of the Pearson correlation matrix $\Gamma$ in
Eq.~(\ref{eq:gamma}) results from the fact that we have projected the
$q$-dimensional space defined by the amino-acids $a=1,\ldots, q$ onto
the subspace spanned by the first $q-1$ dimensions. Alternative
projections lead to modified but equivalent expressions of the
Pearson matrix, cf.~Text S1 (Sec. S1.3). Informally speaking, the
correlation $\Gamma_{ij}(a,b)$ is a measure of comparison of the
empirical covariance $C_{ij}(a,b)$ with the single-site fluctuations
taken independently. Hence, $\Gamma$ is normalized and coincides with
the $(q-1)\times (q-1)$ identity matrix on each site:
$\Gamma_{ii}(a,b)=\delta_{a,b}$.

We further introduce the eigenvalues and eigenvectors ($\mu=1,...,L(q-1)$)
\begin{equation}
\sum_{j=1}^L \sum_{b=1}^{q-1} \Gamma_{ij}(a,b) v_{jb}^\mu 
= \lambda_\mu v_{ia}^\mu\ ,
\label{eq:ev}
\end{equation}
where the eigenvalues are ordered in decreasing order
$\lambda_1\geq\lambda_2\geq\dots\geq\lambda_{L(q-1)}$. The eigenvectors are
chosen to form an ortho-normal basis,
\begin{equation}
\sum_{ia} v_{ia}^\mu v_{ia}^\nu = L\; \delta_{\mu,\nu}\ ,
\end{equation}
for all $\mu,\nu=1,...,L(q-1)$. 
Principal component analysis consists in a partial eigendecomposition
of $\Gamma$, keeping only the eigenmodes contributing most to 
the correlations, {\em i.e.}~with the largest
eigenvalues. All the other eigenvectors are discarded. In this way, 
the directions of maximum covariation of the residues are identified. 

PCA was used by Casari {\it et al.}~\cite{casari95} in the context 
of residue covariation to identify functional sites specific to
subfamilies of a protein family given by a large MSA. To do so, 
the authors diagonalized the comparison matrix, whose elements 
${\cal C}(m,m')$ count the number of identical residues for each 
pair of sequences ($m,m'=1,\ldots M$). Projection of sequences
onto the top eigenvectors of the matrix ${\cal C}$ allows to identify
groups of subfamily-specific co-conserved residues responsible for 
subfamily-specific functional properties, called 
specificity-determining positions (SDP). Up to date, PCA (or the
closely related multiple correspondence analysis) is used in
one of the most efficient tools, called S3det, to detect SDPs
\cite{rausell2010}. PCA was also used in an 
approach introduced by Ranganathan and coworkers 
\cite{lockless99,halabi09}, called statistical coupling analysis 
(SCA). In this approach a modified residue covariance matrix,
$\tilde C^{SCA}$, is introduced : 
\begin{equation}
\tilde C_{ij}^{SCA} (a,b)= w_i^a\; C_{ij}(a,b)\; w_j^b \, 
\end{equation} 
where the weights $w_i^a$ favor positions $i$ and residues $a$ of
high conservation. Amino-acid indices are contracted to
define the effective covariance matrix,
\begin{equation}
\tilde C_{ij}^{SCA} = \sqrt{ \sum_{a,b} \tilde C_{ij}^{SCA}(a,b)^2}\ . 
\end{equation} 
The entries of $\tilde C^{SCA}$ depend on the residue positions $i,j$
only. In a variant of SCA the amino-acid information is directly
contracted at the level of the sequence data. A binary variable is
associated to each site: it is equal to one in sequences carrying the
consensus amino-acid, to zero otherwise \cite{halabi09}. Principal
component analysis can then be applied to the $L$-dimensional $\tilde
C^{SCA}$ matrix, and used to define so-called sectors, 
{\it i.e.}~clusters of evolutionarily correlated sites.

\section*{Results}
\label{sec:hopfield-potts}

To bridge these two approaches -- DCA and PCA -- we introduce the
Hopfield-Potts model for the maximum likelihood modeling of the 
sequence distribution, given the residue frequencies $f_i(a)$ and 
their pairwise correlations $f_{ij}(a,b)$. From a mathematical point 
of view, the model corresponds to a specific class of Potts models, in
which the coupling matrix $e_{ij}(a,b)$ is of low rank $p$ compared to
$L(q-1)$. It therefore offers a natural way to reduce the number of
parameters far below what is required in the mean-field
approximation of \cite{morcos11}. In addition, the solution of the 
Hopfield-Potts inverse problem, {\it i.e.}~the determination of the 
low rank coupling matrix $e$, allows us to establish a direct 
connection with the spectral properties of the Pearson correlation 
matrix $\Gamma$ and thus with PCA.

Here, we first give an overview over the most important
theoretical results for Hopfield-Potts model inference, increasing
levels of detail about the algorithm and its derivation are provided
in Methods and Text S1. Subsequently we discuss in detail
the features of the Hopfield-Potts patterns found in three different
protein families, and finally assess our capacity to detect residue
contacts using sequence information alone in a larger test set of 
protein families.

\subsection*{Inference with the Hopfield-Potts model}

The main idea of this work is that, though the space of sequences is 
$L(q-1)$--dimensional, the number of spatial directions being relevant 
for covariation is much smaller. Such a relevant direction is called 
{\it pattern} in the following, and given by a $L\times q$ matrix
$\boldsymbol\xi=\{\xi_{i} (a)\}$, with $i =1,...,L$ being the site 
indices, and $a=1,...,q$ the amino acids. The {\it log-score} of a 
sequence $(a_1,...,a_L)$ for one pattern $\boldsymbol\xi$ is 
defined as
\begin{equation}\label{defscore}
S(a_1,...,a_L | \boldsymbol\xi )
=\left[ \sum_{i=1}^L \xi_i (a_i) \right] ^{\-2} \ .
\end{equation}
This expression bears a strong similarity with, but also a crucial 
difference to a position-specific scoring matrix (PSSM): As in a 
PSSM, the log-score depends on a sum over position and amino-acid
specific contributions, but its non-linearity (the square
in Eq.~(\ref{defscore})) introduces residue-residue couplings, and thus
is essential to take covariation into account.

In the Hopfield-Potts model, the probability of an amino-acid 
sequence $(a_1,\ldots , a_L)$ depends on the combined log-scores 
along a number $p$ of patterns through 
\begin{equation}\label{ll}
P(a_1,\ldots ,a_L) = \frac 1{\cal Z} \exp \left\{
\frac 1{2L} 
\sum_{\mu=1}^{p_+}S(a_1,\ldots, a_L | \boldsymbol \xi^{+,\mu} ) 
- \frac 1{2L} 
\sum_{\nu=1}^{p_-}S(a_1,\ldots, a_L | \boldsymbol \xi^{-,\nu} )
+ \sum_{i=1}^L h_i(a_i) \right\} \ .
\end{equation} 
Patterns denoted with a $+$-superscript, $\boldsymbol\xi^{+,\mu}$ 
with $\mu=1,...,p_+$, are said to be {\em attractive}, while the patterns 
labeled with a $-$-superscript, $\boldsymbol\xi^{-,\nu}$ for 
$\nu=1,...,p_-$, are called {\em repulsive}. For the probability 
$P(a_1,...,a_L)$ to be large, the log-scores $S(a_1,...,a_L|\boldsymbol \xi)$ 
for attractive patterns must be large, whereas the log-scores for 
repulsive patterns must be small (close to zero). As we will see in
the following, the inclusion of such repulsive patterns is important:
Compared to the mixed model (\ref{ll}), a model with only attractive 
patterns achieves a much smaller likelihood (at each given total number 
of parameters) and a strongly reduced predictivity of residue-residue 
contacts.

It is easy to see that Eq.~(\ref{ll}) corresponds to a specific 
choice of the couplings $e_{ij}(a,b)$ in Eq.~(\ref{eq:potts}), namely 
\begin{equation}
e_{ij}(a,b) = \frac 1L \sum_{\mu=1}^{p_+} \xi_{i}^{+,\mu}(a)\, \xi_{j}^{+,\mu}(b) 
-\frac 1L \sum_{\nu=1}^{p_-} \xi_{i}^{-,\nu}(a)\, \xi_{j}^{-,\nu}(b) \ ,
\label{eq:hopfield}
\end{equation}
where, without loss of generality, the $q^{th}$ component of the 
patterns is set to zero, $\xi_{i}^{+,\mu}(q)=\xi_{i}^{-,\nu}(q)=0$, for 
compatibility with the mean-field approach exposed above. Note that 
the coupling matrix, for linearly independent patterns, has rank 
$p=p_++p_-$, and is defined from $p\,L(q-1)$ pattern components only, 
instead of ${\cal O}(L^2(q-1)^2)$ parameters for the most general case 
of coupling matrices $e_{ij}(a,b)$. When $p=L (q-1)$, {\em i.e.} when 
all the patterns are taken into account, the coupling matrix $e$ has 
full rank, and the Hopfield-Potts model is identical to the Potts 
model used to infer the couplings in DCA in \cite{morcos11}. All 
results of mean-field DCA are thus recovered in this limiting case.

The patterns are to be determined by ML inference, cf.~Methods and 
Text S1 for details. In mean-field approximation, they can be
expressed in terms of the eigenvalues and eigenvectors of the Pearson
correlation matrix $\Gamma$, which were defined in Eq.~(\ref{eq:ev}). 
We find that attractive patterns correspond to the $p_+$ largest 
eigenvalues ($\lambda_1\geq\lambda_2\geq ...\geq\lambda_{p_+}\geq1$),
\begin{equation}
\xi_{i}^{+,\mu} (a)= \left(1-\frac 1{\lambda_\mu} \right)^{1/2}\ 
\tilde v_{ia}^\mu\ , \qquad \mu=1,\ldots ,p_+,
\label{eq:pattern+}
\end{equation}
and repulsive patterns to the $p_-$ smallest eigenvalues 
($\lambda_{L(q-1)}\leq\lambda_{L(q-1)-1}\leq ...\leq\lambda_{L(q-1)+1-p_-}\leq 1$),
\begin{equation}
\xi_{i}^{-,\nu} (a)= \left(\frac 1{\lambda_{L(q-1)+1-\nu} }-1\right)^{1/2}\  
\tilde v_{ia}^{L(q-1)+1-\nu}\ , \qquad  \nu=1,...,p_-,
\label{eq:pattern-}
\end{equation}
where, for all $\mu=1,...,L(q-1)$,
\begin{equation}
\tilde v_{ia}^\mu = \sum_{b=1}^{q-1}
(D_i)^{-1}(a,b)\; v_{ib}^\mu\ .
\label{defvtilde}
\end{equation}
The prefactor $|1-1/\lambda|^{1/2}$ vanishes for $\lambda = 1$. It is 
not surprising that $\lambda=1$ plays a special role, as it coincides 
with the mean of the eigenvalues:
\begin{equation}
\frac 1 {L(q-1)} \sum _{\mu=1}^{L(q-1)} \lambda _\mu = \frac 1 {L(q-1)} 
\sum_{i=1}^L \sum_{a=1}^{q-1} \Gamma _{ii}(a,a) =1 \ .
\end{equation}
In the absence of any covariation between the residues $\Gamma$ 
becomes the identity matrix, and all eigenvalues are unity. Hence all 
patterns vanish, and so does the coupling matrix (\ref{eq:hopfield}). 
The Potts model (\ref{ll}) depends only on the local bias parameters 
$h_i(a)$, and it reduces to a PSSM describing independent sites.

\begin{figure}[!h]
\vspace{1.8cm}
\includegraphics[width=\columnwidth]{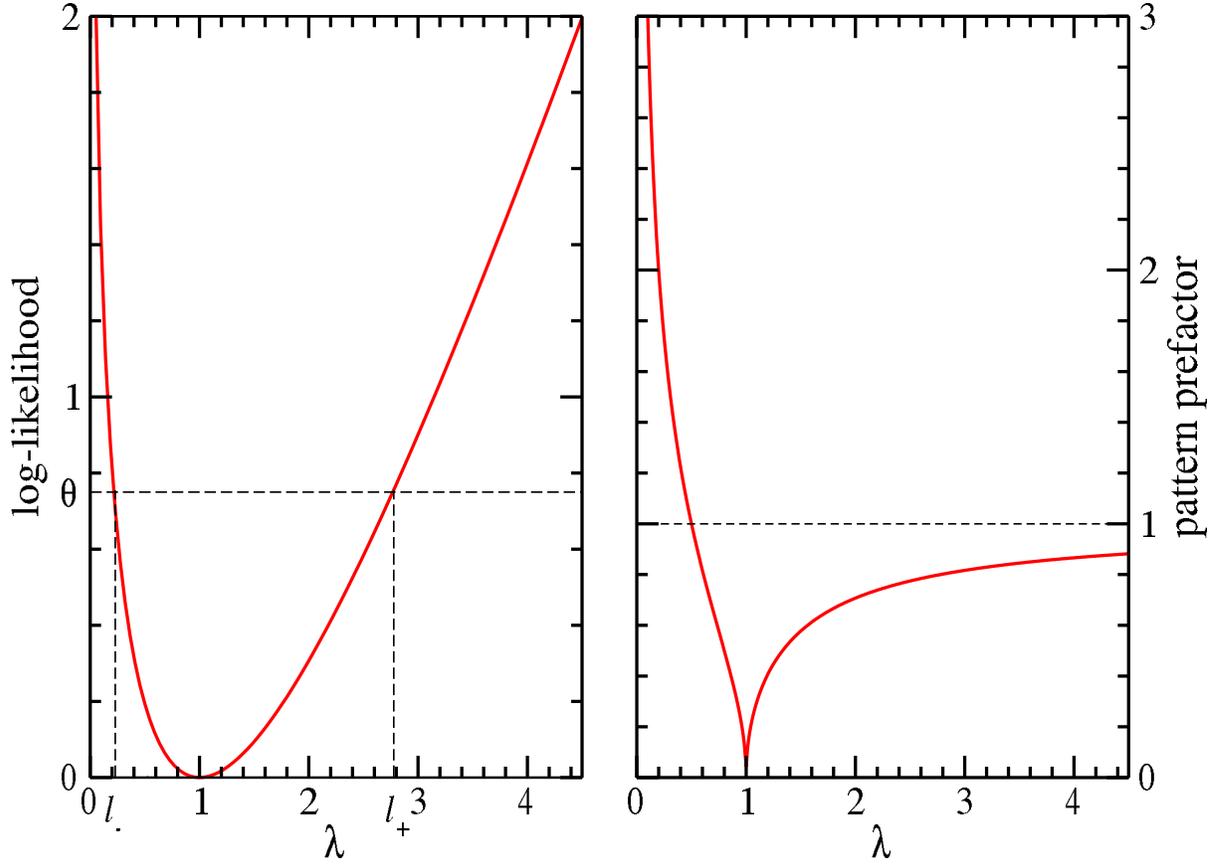}
\vspace{0.8cm}
\caption{{\bf Pattern selection by maximum likelihood and pattern
    prefactors:} {\em (left panel)} Contribution of patterns to the
  log-likelihood (full red line) as a function of the corresponding
  eigenvalues $\lambda$ of the Pearson correlation matrix $\Gamma$. To
  select $p$ patterns, a log-likelihood threshold $\theta$ (dashed
  black line) has to be chosen such that there are exactly $p$
  patterns with $\Delta {\cal L}(\lambda_\mu)>\theta$.  This
  corresponds to eigenvalues in the left and right tail of the
  spectrum of $\Gamma$. {\em (right panel)} Pattern prefactors
  $|1-\frac 1\lambda|^{1/2}$ (full red line) as a function of the
  eigenvalue $\lambda$. Patterns corresponding to $\lambda \simeq 1$
  have essentially vanishing prefactors; patterns associated to large
  $\lambda$ ($\gg 1$) have prefactors smaller than 1 (dashed black
  line), while patterns corresponding to small $\lambda$ ($\ll 1$)
  have unbounded prefactors.  }
\label{fig:ll}
\end{figure}

The eigenvectors of the correlation matrix with large eigenvalues 
$\lambda_\mu \gg 1$ contribute most to the covariation observed in 
the MSA (i.e.~to the matrix $\Gamma$), but they do not contribute 
most to the coupling matrix $e$. In the expression (\ref{eq:hopfield}) 
for this matrix, each pattern carries a prefactor $|1-1/\lambda_{\mu}|$: 
Whereas this prefactor remains smaller than one
for attractive patterns ($\lambda_\mu>1$), it can become very large 
for repulsive patterns ($\lambda_\mu<1$), see Fig.~\ref{fig:ll} 
(right panel). Thus, the contribution of a
repulsive patterns to the $e$ matrix may be much larger than the
contribution of any attractive pattern.

Eqs. (\ref{eq:pattern+}) and (\ref{eq:pattern-}) {\it a priori} 
define $L(q-1)$ different patterns, therefore we need a rule for 
selecting the $p$ 'best', i.e.~most likely patterns. We show in Methods 
that the contribution of a pattern to the model's log-likelihood  
${\cal L}$ defined in Eq.~(\ref{eq:ll_intro}) is a function of the 
associated eigenvalue $\lambda$ only,
\begin{equation}\label{deltas}
\Delta {\cal L} (\lambda) = \frac 12 \bigg(\lambda-1- \log \lambda \bigg)\ .
\end{equation}
As is shown in Fig.~\ref{fig:ll} (left panel), large contributions 
arrive from both
the largest and the smallest eigenvalues, whereas eigenvalues close to
unity contribute little. According to ML inference, we have to select 
the $p$ eigenvalues with largest contributions. To this end, we define 
a threshold value $\theta$ such that there are exactly $p$ patterns 
with larger contributions $\Delta{\cal L}>\theta$ to the log-likelihood;
the $L(q-1)-p$ patterns with smaller $\Delta{\cal L}$ are omitted in 
the expression for the couplings Eq.~(\ref{eq:hopfield}). In accordance
with Fig.~\ref{fig:ll}, we determine thus the two positive real roots
$\ell_\pm$ ($\ell_- < 1 < \ell_+$) of the equation
\begin{equation}
\label{eq:lpm}
\Delta {\cal L} (\ell _\pm) = \theta \ ,
\end{equation}
and include all repulsive patterns with $\lambda_{L(q-1)+1-\nu}<\ell_-$, calling 
their number $p_-$, and all attractive patterns with 
$\lambda_\mu>\ell_+$, denoting their number by $p_+$. The total
number of selected patterns is thus $p=p_-+p_+$.

\subsection*{Features of the Hopfield-Potts patterns}

We have tested the above inference framework in great detail using 
three protein families, with variable values of protein length $L$ 
and sequence number $M$:
\begin{itemize}
\item The {\it Kunitz/Bovine pancreatic trypsin inhibitor} domain
  (PFAM ID PF00014) is a relatively short ($L=53$) and not very
  frequent ($M=2,143$) domain, after reweighting the effective number
  of diverged sequences is $M_{eff}= 1,024$ (cf.~Eq.~(\ref{eq:Meff}) in
  Methods for the definition). Results are compared to the exemplary
  X-ray crystal structure with PDB ID 5pti \cite{wlodawer84}.
\item The bacterial {\it Response regulator} domain (PF00072) is of
  medium length ($L=112$) and very frequent ($M=62,074$).  The
  effective sequence number is $M_{eff}=29,408$. The PDB structure
  used for verification has ID 1nxw \cite{bent04}.
\item The eukaryotic signaling domain {\it Ras} (PF00071) is the
  longest ($L=161$) and has an intermediate size MSA ($M=9,474$),
  leading to $M_{eff}=2,717$. Results are compared to PDB entry 5p21
  \cite{pai90}.
\end{itemize}
In a second step, we have used the 15 protein families studied in 
\cite{sulkowska12} to verify that our findings are not specific to the
three above families, but generalize to other families. A list of the
15 proteins together with the considered PDB structures is provided
in Text S1, Section 4.

To interpret the Hopfield patterns in terms of amino-acid sequences, 
we first report some empirical observations made for the patterns
corresponding to the largest and smallest eigenvalues, {\em i.e.} to 
the most likely attractive and repulsive patterns. We concentrate our 
discussion in the main text on one protein family, the Trypsin 
inhibitor (PF00014). Analogous properties in the other two protein 
families are reported in Text S1.

\begin{figure}[!h]
\vspace{0.8cm}
\includegraphics[width=\columnwidth]{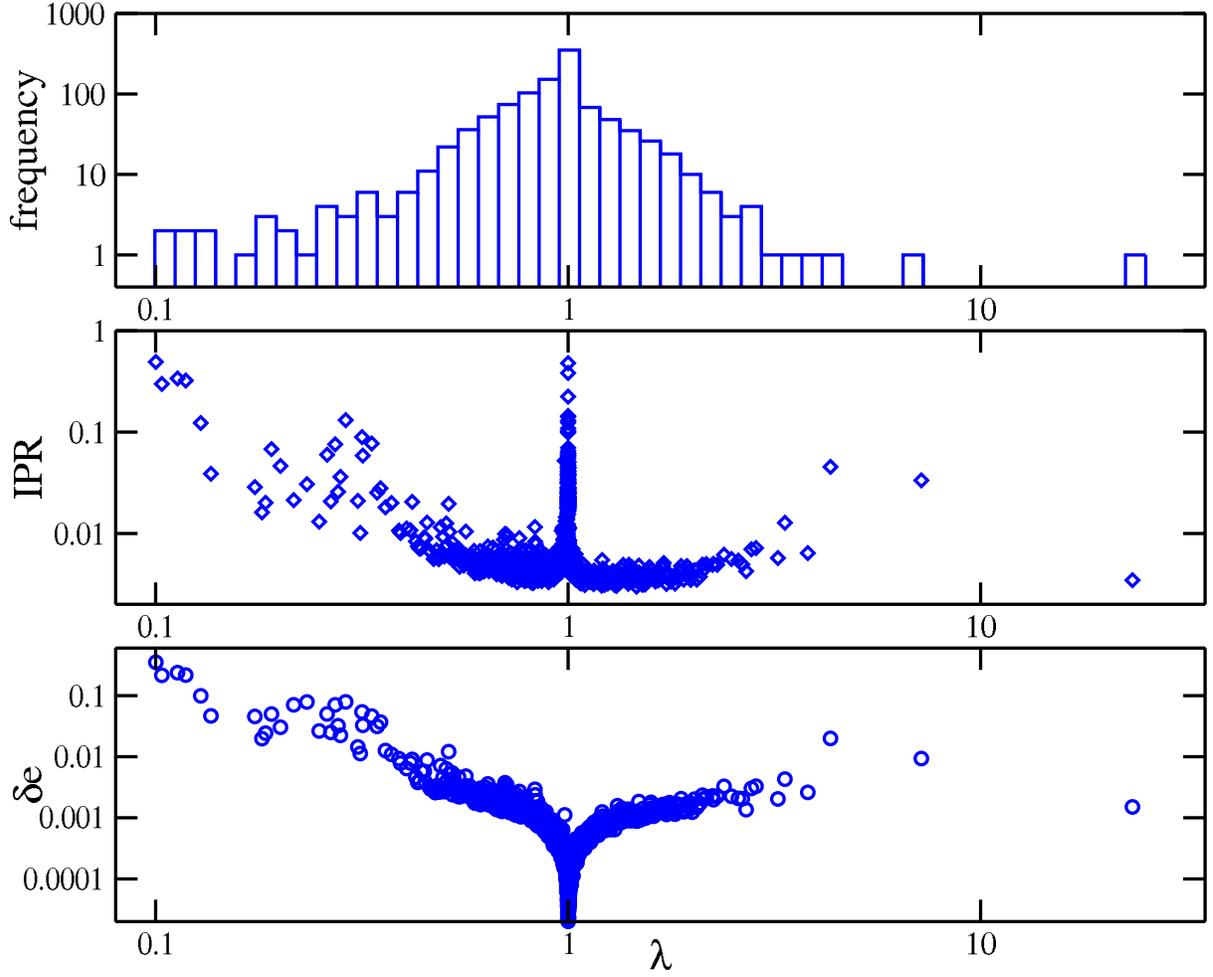}
\vspace{0.8cm}
\caption{{\bf Eigenvalues, localization and contributions to couplings
    for PF00014} (from top to bottom): {\em (top panel)} Spectral
  density as a function of the eigenvalues $\lambda$, note the
  existence of few very large eigenvalues, and a pronounced peak in
  $\lambda=1$.  {\em (middle panel)} Inverse participation ratio of
  the Hopfield patterns as a function of the corresponding eigenvalue
  $\lambda$.  Large IPR characterizes the concentration of a pattern
  to few positions and amino acids.  {\em (bottom panel)} Typical
  contribution $\delta e$ to couplings due to each Hopfield pattern,
  defined in Eq. (\ref{estimatedeltae}), as a function of the
  corresponding eigenvalue $\lambda$. Large contributions are mostly
  found for small eigenvalues, while patterns corresponding to
  $\lambda\simeq 1$ do not contribute to couplings.  }
\label{fig:spectrum}
\end{figure}

The upper panel of Fig.~\ref{fig:spectrum} shows the spectral 
density. It is characterized by a pronounced peak around eigenvalue
1. The smallest eigenvalue is $\lambda^{PF00014}_m\sim 0.1$, the 
largest is $\lambda^{PF00014} _M \sim 23$. Large eigenvalues are
isolated from the bulk of the spectrum, small eigenvalues are not.


To characterize the statistical properties of the patterns we define, 
inspired by localization theory in condensed matter physics, the 
inverse participation ratio (IPR) of a pattern $\boldsymbol\xi$ as
\begin{equation}
\text{IPR}(\boldsymbol\xi) = \frac{\displaystyle{ \sum_{i,a}\xi_{i}(a)^4} } 
{\displaystyle{\big( \sum_{i,a} \xi_{i}(a)^2\big)^2} }\ .
\end{equation}
Possible IPR values range from one for perfectly localized patterns 
(only one single non-zero component) to $1/(L(q-1))$ for a completely
distributed pattern with uniform entries. IPR is therefore used as a 
localization measure for the patterns: Its inverse
$1/\text{IPR}(\boldsymbol \xi)$ is an estimate of the effective number 
$N_{eff}(\boldsymbol\xi)$
of pairs $(i,a)$, on which the pattern has sizable entries $\xi_{i}(a)$. 
The middle panel of Fig.~\ref{fig:spectrum} shows the presence of 
strong localization for repulsive patterns (small eigenvalues) and for
irrelevant patterns (around eigenvalue 1). A much smaller increase in 
the IPR is also observed for part of the large eigenvalues.

What is the typical contribution $\delta e(\boldsymbol\xi)$ of a pattern 
$\boldsymbol\xi$ to the couplings? Pattern $\boldsymbol\xi$ contributes 
$\delta e_{ij}(a,b)=\frac 1L \xi_i(a)\xi_j(b)$ to each coupling. Many 
contributions can be small, and others may be larger. An estimate of 
the magnitude of those relevant contributions can be obtained from the 
sum of the squared contributions normalized by the effective number 
$N_{eff}(\boldsymbol\xi)^2$ of pairs $(i,a),(j,b)$ on which the patterns 
has large entries:
\begin{equation}\label{estimatedeltae}
\delta e(\boldsymbol\xi) = \sqrt{\frac 1{N_{eff}(\boldsymbol\xi)^2} \sum_{i,j,a,b}\big(\delta e_{ij}(a,b) \big)^2} =  \text{IPR}(\boldsymbol\xi)\times
\frac 1L \sum_{i,a} \xi_i(a)^2  \ .
\end{equation}
The lower panel of Fig.~\ref{fig:spectrum} shows the typical contribution 
$\delta e$ of a pattern as a function of its corresponding eigenvalue. 
Patterns with eigenvalues close to 1 have very small norms; they 
essentially do not contribute to the couplings. Highly localized patterns
of large norm result in few and large contributions to the couplings 
($\lambda \ll 1$). Patterns associated to large eigenvalues $\lambda \gg 1$ 
produce many weak contributions to the  couplings. 

\subsubsection*{Repulsive patterns}

In the upper row of Fig.~\ref{fig:patterns} we display the three 
most localized repulsive patterns (smallest, 3rd and 4th smallest
eigenvalues) for the trypsin inhibitor protein (PF00014). All three 
patterns have two very pronounced peaks, corresponding to, say, 
amino-acid $a$ in position $i$ and amino-acid $b$ in position $j$, 
and some smaller minor peaks, resulting in IPR values above 0.3. 
For each pattern, the two major peaks are of opposite sign: 
$\xi_i(a) \simeq - \xi_j(b)$. As a consequence, amino-acid sequences 
carrying amino-acid $a$ in position $i$, but not $b$ in position 
$j$ (as well as sequences carrying $b$ in $j$ but not $a$ in $i$) 
show large log-scores $S \simeq [\xi_i(a)]^2$, 
cf.~Eq.~(\ref{defscore}). Their probability in the Hopfield-Potts 
model, given by (\ref{ll}), will be strongly reduced as compared to 
the probability of sequences carrying either both amino-acids $a$ 
and $b$ in, respectively, positions $i$ and $j$, or none of the two
(scores $S$ close to zero). Hence, we see that repulsive patterns do 
define repulsive directions in the sequence space, which tend to be 
avoided by sequences. A more thorough discussion of the meaning of 
repulsive patterns will be given in the Discussion Section.
  
In all three panels of Fig.~\ref{fig:patterns}, the two large peaks 
have highest value for the amino acid cysteine. Actually, for all of
them, the pairs of peaks identify disulfide bonds, {\it i.e.} 
covariant bonds between two cysteines. They are very 
important for a protein's stability and therefore highly conserved. 
The corresponding repulsive patterns forbid amino-acid configurations 
with a single cysteine in only one out of the two positions. 
Both residues are co-conserved. Note also that the trypsin inhibitor 
has only three disulfide bonds, {\it i.e.} all of them are seen by the most 
localized repulsive patterns. The second eigenvalues, which has a 
slightly smaller IPR, is actually found to be a mixture of two of 
these bonds, {\it i.e.} it is localized over four positions.

The observation of disulfide bonds is specific to the trypsin
inhibitor. In other proteins, also the ones studied in this paper, we
find similarly strong localization of the most repulsive patterns, but
in different amino acid combinations. As an example, the most localized
pattern in the response regulator domain connects a position with
an Asp residue (negatively charged), with another position carrying 
either Lys or Arg (both positively charged), their interaction is
thus coherent with electrostatics. In all observed cases, the 
consequence is a strong statistical coupling of these positions, 
which are typically found in direct contact.

\subsubsection*{Attractive patterns}

The strongest attractive pattern, {\it i.e.} the one corresponding to the 
largest eigenvalue $\lambda_1$, is shown in the leftmost panel of the 
lower row of Fig.~\ref{fig:patterns}. Its IPR is small ($\sim 0.003$), 
implying that it is extended over most of the protein (a  pattern of 
constant entries would have IPR $1/(L(q-1))\simeq 0.001$). As is 
shown in Text S1, strongest entries in $\xi^1_{i}(a)$ 
correspond to conserved residues and these, even if they are 
distributed along the primary sequence, tend to form spatially 
connected and functionally important regions in the folded protein 
({\it e.g.} a binding pocket),
cf.~left panel of Fig.~\ref{fig:sector_contact}. Clearly this
observation is reminiscent of the protein sectors observed in 
\cite{halabi09}, which are found by PCA applied to the before-mentioned 
modified covariance matrix. Note, however, that sectors are extracted 
from more than one principal component.

More characteristic patterns are found for the second and third
eigenvalues. As is shown in Fig.~\ref{fig:patterns}, they show 
strong peaks at the extremities of the sequence, which become higher 
when approaching the first resp. last sequence position. The 
peaks are, for all relevant positions, concentrated
on the gap symbol. These patterns are actually artifacts of the
multiple-sequence alignment: Many sequences start or end with
a stretch of gaps, which may have one out of at least three 
reasons: (1) The protein under consideration does not match the 
full domain definition of PFAM; (2) the local nature of PFAM
alignments has initial and final gaps as algorithmic artifacts, 
a correction would however render the search tools less efficient; 
(3) in sequence alignment algorithms, the extension of an 
existing gap is less expensive than opening a new gap.
The attractive nature of these two patterns, and the equal sign 
of the peaks, imply that gaps in equilibrium configurations of 
the Hopfield-Potts model frequently come in stretches, and 
not as isolated symbols. The finding that there are two patterns 
with this characteristic can be traced back to the fact that
each sequence has two ends, and these behave independently with
respect to alignment gaps.

\subsubsection*{Theoretical results for localization in the 
limit case of strong conservation}

The main features of the empirically observed spectral and 
localization properties of Fig.~\ref{fig:spectrum} can be found 
back in the limiting case of completely conserved sequences, which 
is amenable to an exact mathematical treatment. To this end, we 
consider $L$ perfectly conserved sites, {\it   i.e.}~a MSA made 
from the repetition of a unique sequence. As is shown in Text S1, Section 2, the corresponding Pearson correlation 
matrix $\Gamma$ has only three different eigenvalues:
\begin{itemize}
\item a large and non-degenerate eigenvalue, $\lambda _+$, which 
is a function of $q$ and $L$ (and of the pseudocount used to 
treat the data, see Methods), whose corresponding eigenvector 
is extended; 
\item a small and $(L-1)$-fold degenerate eigenvalue, 
$\lambda_-= (L-\lambda_+)/(L-1)$. The corresponding eigenspace 
is spanned by vectors which are perfectly localized in pairs of 
sites, with components of opposite signs;  
\item the eigenvalue $\lambda=1$, which is $L(q-2)$-fold 
degenerate. The eigenspace is spanned by vectors, which are 
localized over single sites. 
\end{itemize}  
For a realistic MSA, {\em i.e.} without perfect conservation, 
degeneracies will disappear, but the features found above remain 
qualitatively correct. In particular, we find in real data a 
pronounced peak of eigenvalues around 1, corresponding to localized 
eigenmodes (Fig.~\ref{fig:spectrum}) . In addition, low-eigenvalue 
modes are found to be strongly localized, and the the order of 
magnitude of $\lambda_- \simeq 0.09$ is in good agreement with the 
smallest eigenvalues, $\simeq 0.1$, reported for the three  analyzed 
domain families. Finally, the largest eigenmodes are largely 
extended, as found in the limit case above. Note that the eigenvalues 
found in the protein spectra, {\em e.g.} $\lambda_1\simeq  23$ for 
PF00014, are however smaller than in the limit case, 
$\lambda_+ \simeq 48$, due to only partial conservation in the 
real MSA.

\subsection*{Residue-residue contact prediction with the Hopfield-Potts model}
\label{sec:contact}

\begin{figure}[!h]
\vspace{0.8cm}
\includegraphics[width=\columnwidth]{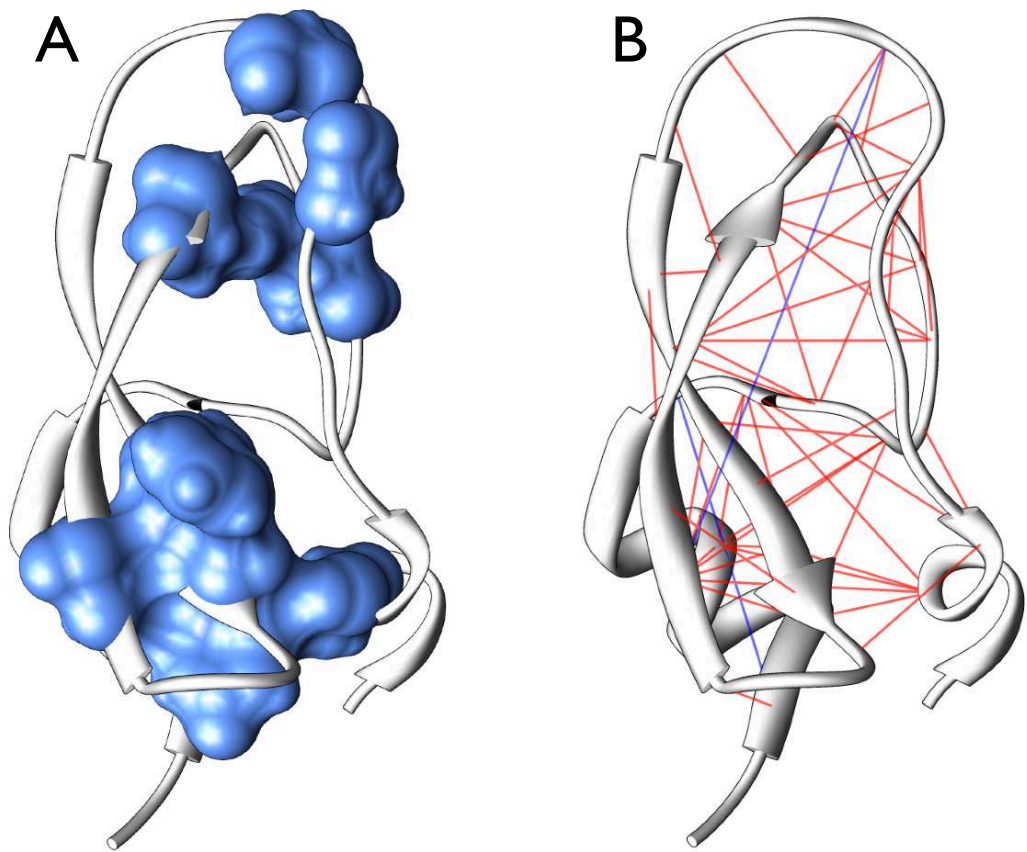}
\vspace{0.8cm}
\caption{{\bf The principal component and predicted contacts visualized 
    on the 3D structure of the trypsin inhibitor protein domain PF00014.} 
  {\em (A)} The 10 positions (residue ID 5,12,14,22,23,30,35,40,51,55) of 
  largest entries in the most attractive   Hopfield pattern (largest 
  eigenvalue of $\Gamma$, corresponding to 
  the principal component) are shown in blue, they correspond also to 
  very conserved sites. Note that, while they are distant along 
  the protein backbone, they cluster into spatially connected components
  in the folded protein. {\em (B)} The 50 residue pairs with strongest 
  couplings (ranked according to the Frobenius norms Eq.~(\ref{corr_norm}), 
  with at least 5 positions separation along the backbone, are connected 
  by lines.  Only two out of these pairs are not in contact (blue links), 
  all other 48 are thus true-positive contact predictions (red links). 
  Many contacts link pairs of not conserved positions. Note that links are 
  drawn between C-alpha atoms, whereas contacts are defined via minimal 
  all-atom distances, making some red lines to appear rather long even if
  corresponding to native contacts.}
\label{fig:sector_contact}
\end{figure}

The most important feature of DCA is its ability to predict pairs of
residues, which are distantly positioned in the sequence, but which
form native contacts in the protein's tertiary structure, cf.~the 
right panel of Fig.~\ref{fig:sector_contact}. Here, our
contact prediction is based on the sampling-corrected Frobenius norm
of the $(q-1)$--dimensional statistical coupling matrices $e_{ij}$,
cf.~Methods, which in \cite{ekeberg12} has been shown to outperform
the direct-information measure used in \cite{weigt09}. This measure
assigns a single scalar value for the strength of the direct coupling
between two residue positions. 

\begin{figure}[!h]
\vspace{2cm}
\includegraphics[width=.9\columnwidth]{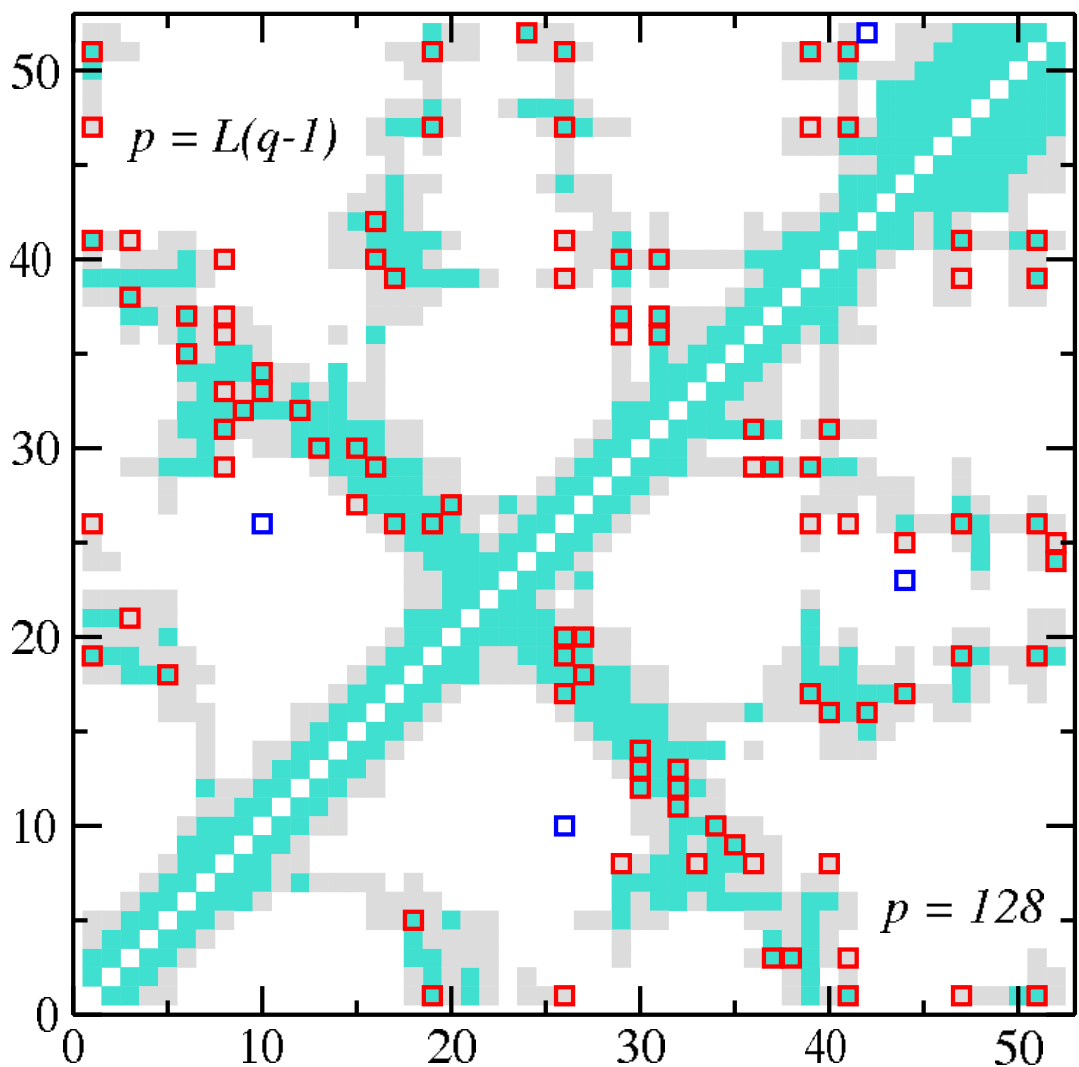}
\vspace{1.8cm}
\caption{{\bf Contact map for the PF00014 family.} Filled squares 
represent the native contact map on the 3D fold (PDB 5pti, with
turquoise squares signaling all-atom distances below 5{\AA}, and 
grey ones distances between 5{\AA} and 8{\AA}). The 50 top predicted 
contacts with minimal separation of 5 positions along the sequence 
($|i-j|\geq 5$) are shown with empty squares: true-positive predictions 
(distance $<8${\AA})
are colored in red, and false-positive predictions in blue. Predictions 
are made with the Hopfield-Potts model  with $p=128$ patterns (bottom 
right corner) and with $p=L(q-1)=1060$ patterns (DCA, top left corner). 
For both values of $p$ there are 48 true-positive and 2 false-positive 
predictions.}
\label{fig:contact_map}
\end{figure}

\begin{figure}[!h]
\vspace{0.8cm}
\includegraphics[width=\columnwidth]{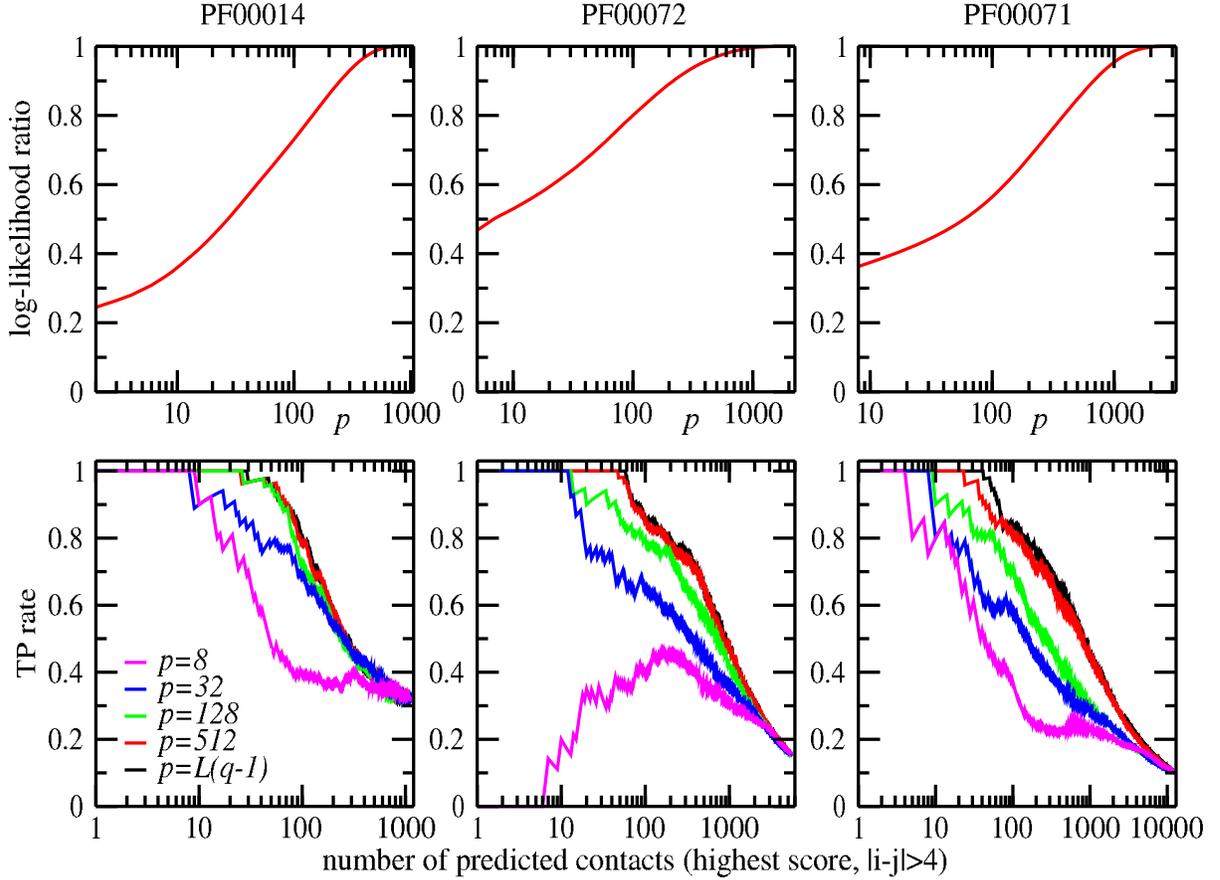}
\vspace{0.8cm}
\caption{{\bf Contact predictions for the three considered protein
  families.} The upper panels show the fraction of the interaction-based
  contribution to the log-likelihood of the model
given the MSA, defined as the ratio of the log-likelihood with   
$p$ selected patterns over the maximal log-likelihood obtained by including all 
$L(q-1)$ patterns, as a function of the number $p$
  of selected patterns, it reaches one for $p=(q-1)L$ corresponding to the
Potts model used in DCA. The lower
  panels show the TP rates as a function of the predicted residue
  contacts, for various numbers $p$ of selected patterns, where
  selection was done using the maximum-likelihood criterion.
$p=(q-1)L$ gives the contact predictions obtained by DCA approach. Only non-trivial contacts between sites $i,j$ such that $|i-j|>4$ are considered in the calculation of the TP rate.}
\label{fig:protein}
\end{figure}

The contact map predicted from the 50 strongest direct couplings 
for the PF00014 family is compared to the native contact map in 
Fig.~\ref{fig:contact_map}. In accordance with \cite{morcos11}, 
a residue pairs is considered to be a true positive prediction
if its minimal atom distance is below 8\AA~in the before 
mentioned exemplary protein crystal structures. This relatively
large cutoff was chosen since DCA was found to extract a bimodal
signal with pairs in the range below 5{\AA} (turquoise in
Fig.~\ref{fig:contact_map}) and others with 7-8{\AA} (grey in 
Fig.~\ref{fig:contact_map}); both peaks contain valuable information 
if compared to typical distances above 20{\AA} for randomly chosen 
residue pairs. 
To include only non-trivial contacts, we require also a minimum separation 
$|i-j|>4$ of at least 5 residues along the protein sequence. Remarkably
the quality of the predicted contact map with the Hopfield-Potts model 
with $p=128$ patterns is essentially the same as with DCA, corresponding 
to $p=L(q-1)=1060$ patterns. In both cases predicted contacts spread 
rather uniformly over the native contact map, and 96\% of the predicted 
contacts are true positives.  This result is corroborated by the lower 
panels of Fig.~\ref{fig:protein}, which show, for various values of the 
number $p$ of patterns, the performance in terms of contact predictions 
for the three families studied here. The plots show the fraction of 
true-positives (TP), {\em i.e.} of native distances below 8\AA, in 
between the $x$ pairs of highest couplings, as a function of $x$ 
\cite{morcos11}. 

The three upper panels in Fig.~\ref{fig:protein} show the ratio
between the selected pattern contributions to the log-likelihood,
$\sum_{\{\mu| \lambda_\mu \notin [\ell_- , \ell_+ ]\}} \Delta {\cal L}
(\lambda_\mu )$, and its maximal value obtained by including all
$L(q-1)$ patterns, $\sum_{\mu=1}^{L(q-1)} \Delta {\cal
  L}(\lambda_\mu)$.  A large fraction of patterns can be omitted
without any substantial loss in log-likelihood, but with a substantially
smaller number of parameters. It is worth noting that, in 
Fig.~\ref{fig:protein}, we do not find
any systematic benefit of excluding patterns for the contact
prediction, but the predictive power decreases initially only very
slowly with decreasing pattern numbers $p$.  For all three proteins,
even with $\sim 128$ patterns, very good contact predictions can be
achieved, which are comparable to the ones with $L(q-1)=1060-3220$ patterns 
using the full DCA inference scheme of \cite{morcos11}. Almost 
perfect performance is reached, when the contribution of selected 
patterns to the log-likelihood is only at $60-80\%$ of its maximal
value. This could be expected from the fact that patterns
corresponding to eigenvalues close to unity hardly contribute to the 
couplings, cf.~lower panel in Fig.~\ref{fig:spectrum}.

\begin{figure}[!h]
\vspace{0.8cm}
\includegraphics[width=\columnwidth]{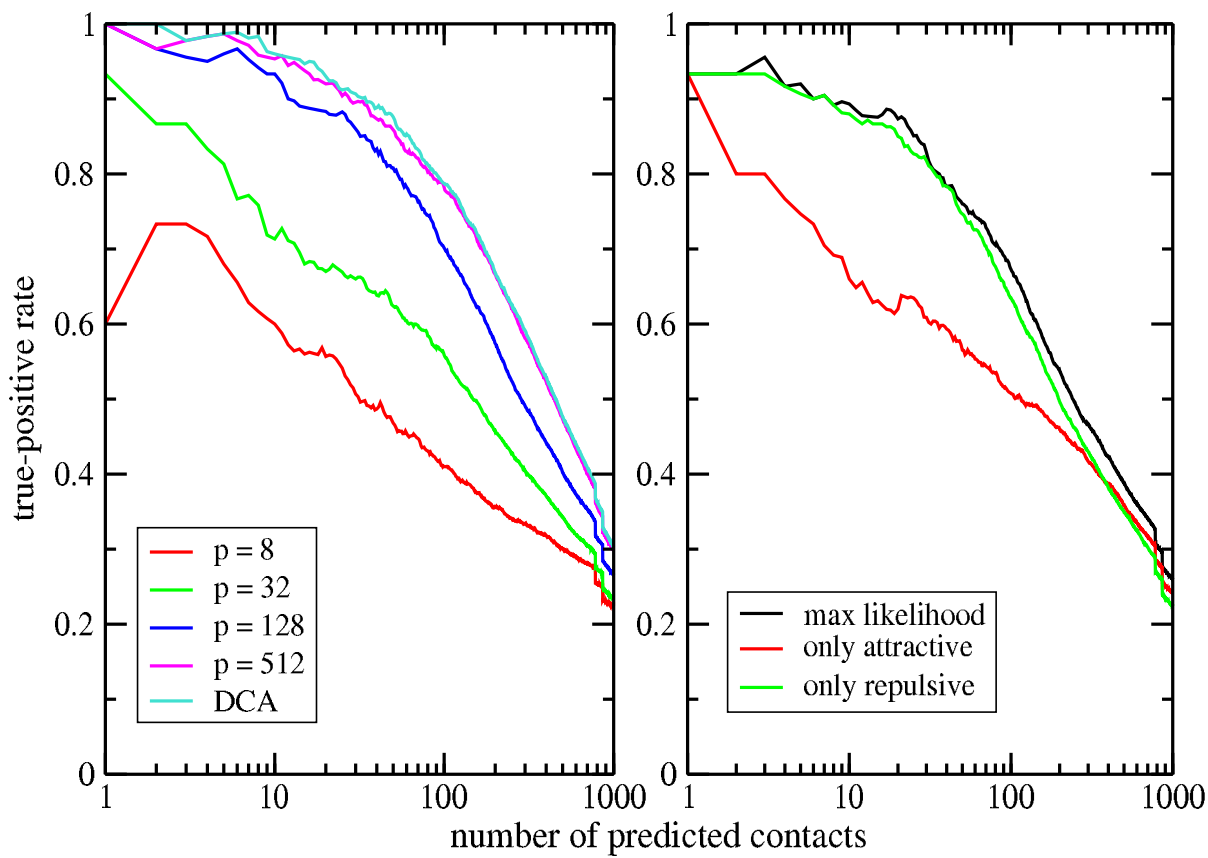}
\vspace{0.8cm}
\caption{{\bf Contact predictions across 15 protein families.} 
  {\it (left panel)} TP rates for the contact prediction 
  with variable numbers $p$ of Hopfield-Potts patterns, averaged over
  15 distinct protein families.
  {\it (right panel)} TP rates for the contact 
  prediction using only the repulsive (green line) resp. attractive 
  (red line) patterns, which are contained in the $p=100$ most likely
  patterns (black line), averaged over 15 protein families. It becomes 
  obvious that the contact prediction remains almost unchanged when
  only the subset of repulsive patterns is used, whereas it drops
  substantially by keeping only attractive patterns. 
}
\label{fig:p100}
\end{figure}

These findings are not restricted to the three test proteins, as is
confirmed by the left panel of Fig.~\ref{fig:p100}. In this figure,
we average the TP rates for $p=8,32,128,512$ and $L(q-1)$ ({\it i.e.} full
mean-field DCA) for the 15 proteins studied in \cite{sulkowska12}, 
which had been selected for their diversity in protein length and fold
type. Further more, the discussion of the localization properties of 
repulsive patterns is corroborated by the results reported in 
Fig.~\ref{fig:p100}, right panel. It compares the performance of the
Hopfield-Potts model to predict residue-residue contacts, for the
three cases where $p=100$ patterns are selected either according to the
maximum likelihood criterion (patterns for eigenvalues $\lambda<\ell_-$ 
and for $\lambda>\ell_+$), or where only the strongest
attractive ($\lambda>\ell_+$) or only the strongest repulsive
($\lambda<\ell_-$) patterns are taken into account. It becomes
evident that repulsive patterns provide more accurate contact 
information, TP rates are almost unchanged between the
curve of the $p=100$ most likely patterns, and the smaller subset
of repulsive patterns. On the contrary, TP rates for contact 
prediction are strongly reduced when considering only attractive 
patterns, {\it i.e.}~in the case corresponding most closely
to PCA. This finding illustrates one of the most significative
differences between DCA and PCA: Contact information is provided by
the eigenvectors of the Pearson correlation matrix $\Gamma$ in the 
lower tail of the spectrum.

As is discussed in the previous section, patterns with the largest
contribution to the log-likelihood are dominated by (and localized in)
conserved sites. Attractive patterns favor these sites to jointly
assume their conserved values, whereas repulsive patterns avoid
configurations where, in pairs of co-conserved sites, only one
variable assumes its conserved value, but not the other one. However,
we have also seen that an accurate contact prediction requires at
least $\sim 100$ patterns, {\it i.e.}~it goes well beyond the patterns
given by strongly conserved sites. In Fig.~\ref{fig:sector_contact} we
show, for the exemplary case of the Trypsin inhibitor, both the 10
sites of highest entry in the most attractive pattern $\boldsymbol\xi^{+,1}$
(corresponding to conserved sites), and the first 50 predicted
intra-protein contacts using the full mean-field DCA scheme (results
for $p=512$ are almost identical). It appears that many of the
correctly predicted contacts are not included in the set of the most
conserved sites. From a mathematical point of view, this is
understandable - only variable sites may show covariation. From
a biological point of view, this is very interesting, since it shows
that highly variable residue in proteins are not necessarily
functionally unimportant in a protein family, but they may undergo
strong coevolution with other sites, and thus be very important for
the structural stability of the protein, cf.~also the Supporting 
Fig.~S5 where the degree of conservation \cite{consurf} 
is depicted for the highest-ranking DCA predicted contacts.
In this figure we show that residues included in predicted contacts
are found for all levels of conservation. It has, however, to be
mentioned that in the considered MSA, there are no 100\% conserved
residues, the latter would not show any covariation. A small level of
variability is therefore crucial for our approach.

A remark is necessary concerning the right panel of
Fig.~\ref{fig:sector_contact}: Whereas conserved sites (which carry
also the largest entries of the pattern with maximum eigenvalue) are
collected in one or two spatially connected regions in the studied
proteins, this is not necessarily true for all proteins. In particular
complex domains with multiple functions and/or multiple conformations
may show much more involved patterns. It is, however, beyond the scope
of this paper to shed light onto the details of the biological
interpretation of the principal components of $\Gamma$.

\begin{figure}[!h]
\vspace{0.8cm}
\includegraphics[width=\columnwidth]{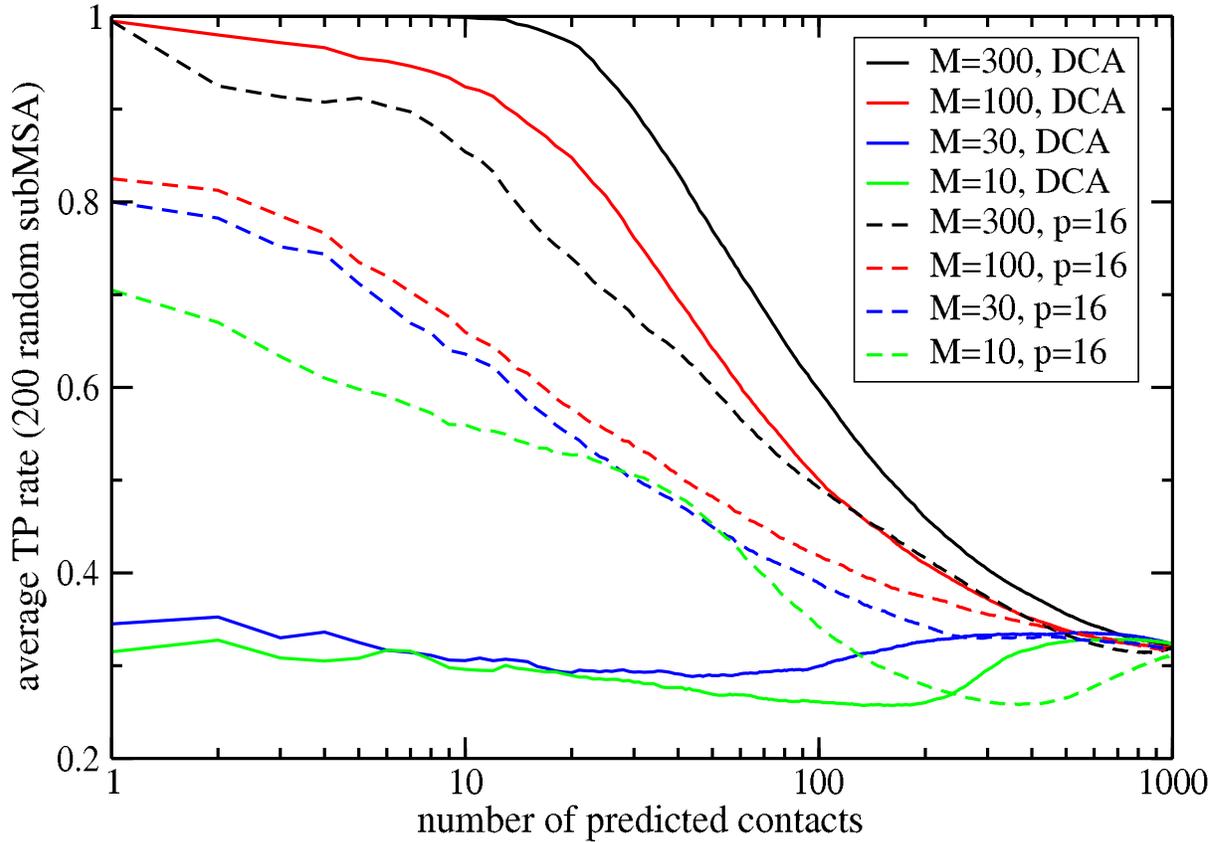}
\vspace{0.8cm}
\caption{ {\bf Noise reduction due to pattern selection in reduced
data sets.} {\it (full lines)} TP rates of mean-field DCA for sub-MSAs 
of family PF00014 with $M=10,30,100,300$ sequences; each curve is 
averaged over 200 randomly selected sub-alignments. Whereas for $M=100$
and $M=300$ the accuracy of the first predictions is close to one,
mean-field DCA does not extract any reasonable signal for $M=10$ and
$M=30$. {\it (dashed lines)} The same sub-MSA are analyzed with the
Hopfield-Potts model using $p=16$ patterns (maximum-likelihood selection).
Whereas this selection reduces the accuracy for $M\geq 100$, it results
in increased TP rates for $M\leq 30$. Dimensional reduction by pattern
selection has lead to an efficient noise reduction.
}
\label{fig:reducedMSA}
\end{figure}

In which cases does the dimensional reduction achieved
by selecting only a relatively small number of patterns provide an
actual advantage over the standard mean-field DCA approach with
$p=L(q-1)$ patterns? We have seen that for relatively large MSAs,
where DCA gives very accurate results, the approach presented here
achieves a strong dimensional reduction almost without loss in 
predictive power, but it did not improve the contact map prediction,
cf.~Figs.~\ref{fig:contact_map} and \ref{fig:protein}. However, when
we reduce the number of sequences in the MSA, DCA undergoes a strong
reduction in accuracy of prediction, see the full lines in 
Fig.~\ref{fig:reducedMSA} where DCA is applied to sub-alignments of
the PF00014 domain family. Repeating the same experiment with a finite
number of patterns ($p=16$ in Fig.~\ref{fig:reducedMSA}), the MSA-size
dependence is strongly reduced. For very small alignments of only 10-30
sequences, the Hopfield-Potts model is still able to extract contacts
with an astonishing TP rate of 70-80\%, whereas DCA produces almost
random results (TP rate ca. 30\%). The success of the Hopfield-Potts approach for small MSA is not specific to the PF00014 domain, and holds for other protein families, see Fig.~S15 in Text S1. Hopfield-Potts patterns are therefore
an efficient means to reduce overfitting effects found in DCA,
and to improve the signal-to-noise ratio.

\section*{Discussion}

In this paper we have proposed a method to analyze the correlation
matrix of residue occurrences across multiple-sequence alignments
of homologous proteins, based on the inverse Hopfield-Potts model. Our
approach offers a natural interpolation between the spectral analysis
of the correlation matrix, carried out in principal component analysis 
(PCA), and maximum entropy approaches which aim at reproducing those 
correlations within a global statistical model ({\it e.g.} DCA). The inverse 
Hopfield-Potts model requires to infer ``directions'' of particular
importance in the sequence space, called patterns: The distribution of
sequences belonging to a protein family tends to accumulate along
attractive patterns (related to eigenmodes of the correlation matrix
with large eigenvalues) and to get depleted around repulsive patterns
(related to the low-eigenvalue modes). These patterns have some
similarity with position-specific scoring matrices frequently used in
the statistical modeling of sequences, but in contrast to the
independence of different positions in PSSM, Hopfield-Potts patterns 
account for inter-position couplings, as needed for coevolutionary 
analysis.

Contrary to principal component
analysis, which discards low-eigenvalue modes, we have shown that
repulsive patterns are essential to characterize the sequence
distribution, and in particular to detect structural properties
(residue-residue contacts) of proteins from sequence data. In addition, 
we have shown how to infer not only the values of the patterns but 
also their statistical relevance from the sequence data. To do so, 
we have calculated the contribution of each pattern to the total 
likelihood of the Hopfield-Potts model given the data, establishing 
thus a clear criterion for pattern selection. The results of the
application of the inverse Hopfield-Potts model to real sequence data
confirm that most eigenmodes (with eigenvalues close to unity) can be
discarded without affecting considerably the contact prediction (see 
Fig.~\ref{fig:contact_map} and Fig.~\ref{fig:protein}). This
makes our approach much less parameter-intensive that the full direct
coupling analysis DCA. We have found empirically that it is
sufficient to take into account the patterns contributing to $\sim
60-80\%$ of the log-likelihood to achieve a very good contact map 
prediction in the case of large multiple-sequence alignments. In the 
case of reduced MSA size, we found that the dimensional reduction
due selecting only the most likely patterns improves the signal-to-noise
ratio of the inferred model, and therefore reaches a better contact
prediction than mean-field DCA, down to very small numbers of sequences, 
see Fig.~\ref{fig:reducedMSA} and Fig.~S15 in Text S1. Moreover the Hopfield-Potts approach
can be very advantageous in terms of computational time. While DCA requires the inversion of
the correlation matrix, which takes $O(L^3(q-1)^3)$ time, computing the $p$ patterns (corresponding to the largest and smallest eigenvalues) 
can be done in $O(pL^2(q-1)^2)$ time only. The reduction in computational time can thus be very important for large proteins.

\begin{figure}[!h]
\vspace{0.8cm}
\includegraphics[width=\columnwidth]{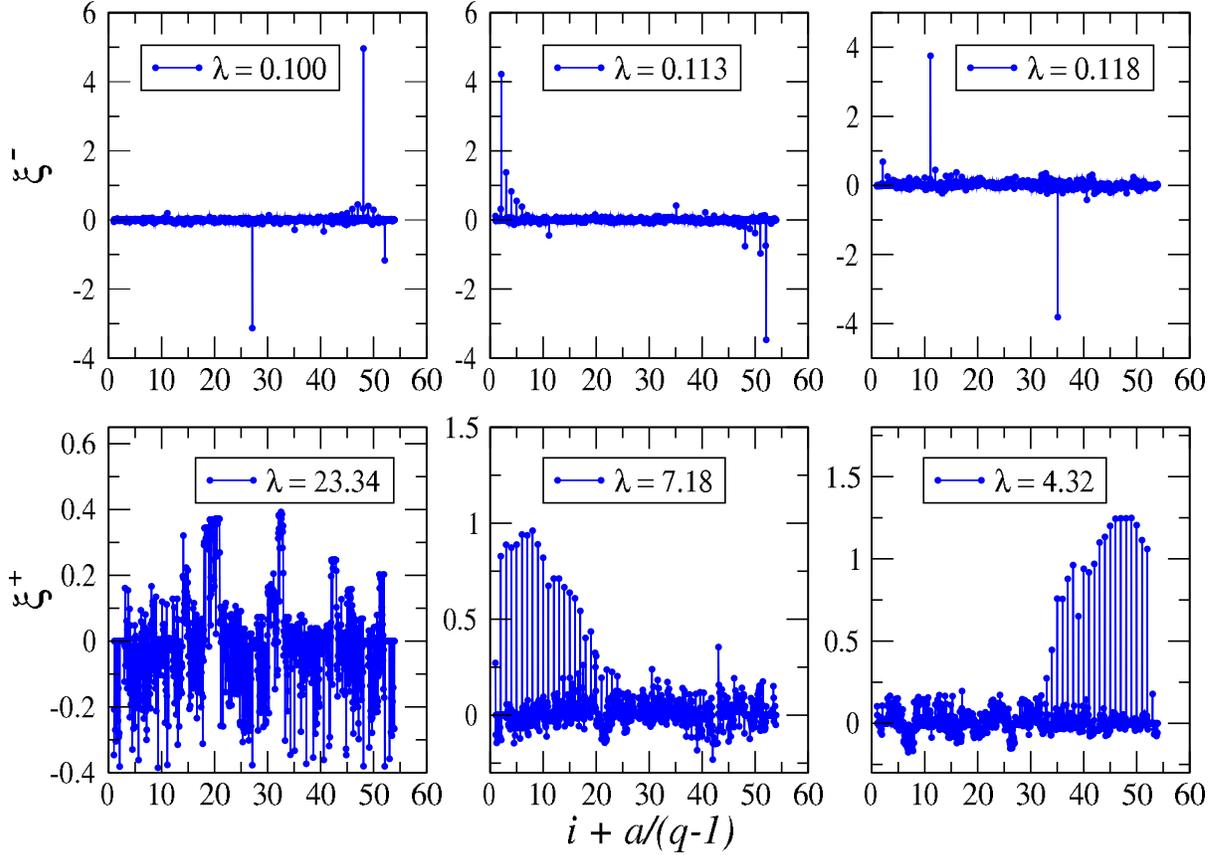}
\vspace{0.8cm}
\caption{{\bf Attractive and repulsive patterns for PF00014:} {\em
    (upper panels)} The most localized repulsive patterns 
  (corresponding to the first, third and fourth smallest eigenvalues
  and inverse participation ratios $0.49, 0.34, 0.32$ respectively) 
  are strongly concentrated in pairs of positions. 
  {\em (lower panels)} The most attractive patterns (corresponding to 
  the three largest eigenvalues); the top pattern is extended, with 
  inverse participation ratio $0.003$, while the second and third
  patterns,with inverse participation ratios $0.033, 0.045$
  respectively, have essentially non-zero components over the gap
  symbols only which accumulate on the edges of the sequence. Note 
  the $x$-coordinates $i+a/(q-1)$; its integer part is the site index, 
  $i$, and the fractional part multiplied by $q-1$ is the residue value, 
  $a$. }
\label{fig:patterns}
\end{figure}

We have also studied the position-specific nature of patterns, taking
inspiration from localization theory in condensed matter physics and
random matrix theory (Fig.~\ref{fig:patterns} and Figs.~S8 and S12 in Text S1). Briefly speaking, a pattern is said to be
localized if it is concentrated on a few sites of the sequence, and
extended (over the sequence) otherwise.  We have found that the
principal attractive pattern (corresponding to the largest eigenvalue)
is extended, with entries of largest absolute value in the most
conserved sites (Figs.~S3, S4, S9 \& S13 in Text S1). Other
strongly attractive patterns can be explained from the presence of
extended gaps in the alignment, mostly found at the beginning or at
the end of sequences. The other patterns of large likelihood
contributions are repulsive, {\em i.e.} they correspond to small
eigenvalues, usually discarded by principal component
analysis. Interestingly, these patterns appear to be strongly
localized, that is, strongly concentrated in very few positions, which
despite their separation along the sequence are found in close contact
in the 3D protein structure. To give an example, in the Trypsin
inhibitor protein, they are localized in position pairs carrying
Cysteine, and being linked by disulfide bonds. Other amino-acid
combinations were also found in the other protein families studied
here, e.g.~patterns connecting residues of opposite electrical charge. 
Taking into account only a number
$p$ of such repulsive patterns results in a predicted contact map of
comparable quality to the one using maximum-likelihood selection,
whereas the same number $p$ of attractive patterns performs
substantially worse (Fig.~\ref{fig:p100} and Fig.~S7 \& S11in Text S1). The dimensional reduction of the
Hopfield-Potts model compared to the Potts model (used in standard
DCA) is thus even more increased as many relevant patterns are
localized and contain only a few (substantially) non-zero components.
As a consequence the couplings found with the Hopfield-Potts model are 
sparser than their DCA counterparts (Fig.~S6 in Text S1).

It is important to stress that also distinct patterns, whether attractive 
or repulsive, can have large components on the same sites and residues.
A general finding, supported by a theoretical analysis in the Results 
section, is that the more repulsive patterns are, the stronger
they are localized, and the more conserved are the residues supporting 
them. Highly conserved sites therefore appear both in the most attractive 
pattern and, when covarying with other residues, in a few localized and 
repulsive patterns reflecting those covariations. As the number of 
patterns to be included to reach an accurate contact prediction is a 
few hundreds for the protein families considered here, the largest 
components of the weakly repulsive patterns, {\em i.e.} with the 
eigenvalues smaller than, but close to the threshold $\theta$, 
correspond to weakly conserved residues. In consequence many predicted 
contacts connect low-conservation residues. This statement is apparent 
from Fig.~\ref{fig:sector_contact} and Figs.~S10
and S14 in Text S1, which compare the sets of conserved sites and the pairs of 
residues predicted to be in contact by our analysis.

Why are repulsive patterns so successful in identifying contacts, in
difference to attractive patterns? To answer this question, consider
the simple case of a pattern $\boldsymbol\xi$  localized in two 
residues only, say it should prefer the co-occurrence of amino-acid 
$a$ in position $i$, and of amino acid $b$ in position $j$. We further 
assume that the two non-zero components $\xi _{i}(a)$ 
and $\xi_{j}(b)$ have the same amplitude and differ only by sign, 
{\em i.e.} $\xi_{i}(a)=-\xi_{j}(b)$. Now we consider a sequence of 
amino-acids $(a_1,\ldots , a_L)$ and ask whether it will have a large 
log-score $S$ for pattern $\boldsymbol\xi$, see Eq.~(\ref{defscore}). 
The outcome is given in the third column of Table 1. 
The log-score therefore corresponds to a {\tt XOR} (exclusive or) 
between the presence of the two amino-acids $a$ and $b$ on their 
respective positions $i$ and
$j$ in the sequence. If the pattern were attractive (cf. fourth
column), it would favor sequences where exactly one of the two
specified amino-acids is present. For a repulsive pattern (cf.~fifth 
column), low log-score sequences are favored, {\it i.e.}~either both 
$a$ and $b$ are present in positions $i$ and $j$, or none of the two.

In case we assumed equal sign components, {\em i.e.} $\xi
_{i}(a)=\xi_{j}(b)$, we would have found Table 2.
This choice is poor in terms of enforcing covariation in the sequence:
An attractive (resp. repulsive) pattern strongly  favors (resp. disfavors) 
the simultaneous presence of amino acids $a$ and $b$ in positions $i$ 
and $j$, but the likelihood is monotonous in the number of correctly 
present amino acids.

As a conclusion, we find that
strong covariation can be efficiently enforced only by a repulsive
pattern with opposite components (fifth column in Table~1). 
The acceptance of the (NO,NO) configuration is desirable, 
too: It signals the possibility of compensatory mutations, {\it i.e.} 
favorable double mutations changing both $a$ and $b$ in positions 
$i$ and $j$ to alternative amino acids. It is easy to generalize 
the above patterns to patterns having more than one favored amino-acid 
combination, {\it e.g.}~favored pairings $(a,b)$ and $(c,d)$ can be
enforced by a repulsive pattern with 
$\xi_{i}(a)=-\xi_{i}(c)=-\xi_{j}(b)=\xi_{j}(d)$. 
This theoretical argument explains why localized repulsive patterns 
critically encode for covariation. Remarkably the condition that the 
few, large components of repulsive patterns should sum up to zero 
agrees well with our findings in real MSAs, cf~Fig.~\ref{fig:patterns} 
and Figs.~S8 and S12 in Text S1. Furthermore, it would be 
interesting to better understand the relationship between such localized 
patterns and specificity-determining positions \cite{casari95,rausell2010}: 
SDP are co-conserved in subfamilies of the full MSA, but vary from one 
family to another. The most repulsive patterns are localized in residues,
which are strongly conserved throughout the full alignment. We have also
used S3det \cite{rausell2010} to predict SDPs and to compare them to
our 30 highest-scoring contact predictions, and we have not observed any
particular signal. It would be interesting to extend the Hopfield-Potts
approach to subfamilies and to investigate, if SDPs correspond to
repulsive patterns in these subfamilies.

Last but not least, let us emphasize the importance of the prefactor 
$|1-\frac 1\lambda|^{1/2}$ of the pattern, cf.~Eqs.~(\ref{eq:pattern+}) 
and (\ref{eq:pattern-}), where $\lambda$ is the eigenvalue corresponding 
to the pattern. While this factor is at most equal to $1$ for attractive 
patterns, it can take arbitrarily large values for repulsive patterns (Fig.~\ref{fig:ll}, right panel). 
Moreover, repulsive patterns can be highly localized: they strongly contribute to a few 
couplings $e_{ij}(a,b)$, {\em e.g.}~to one coupling between a single
pair of positions $i$ and $j$ for patterns perfectly localized in two 
sites only (cf.~Fig.~\ref{fig:spectrum}, lower panel, and Fig.~\ref{fig:patterns}). Consequently those contributions are of particular importance 
in the ranking of couplings, which our contact prediction is based on. 
On the contrary, attractive patterns, even with sizeable norms, produce 
many weaker contributions to the couplings (cf.~Fig.~\ref{fig:spectrum}, lower panel), and do not alter their 
relative rankings a much as repulsive patterns do. This explains why 
contact prediction based on repulsive patterns only is much more 
efficient than when based on attractive patterns only 
(cf.~Fig.~\ref{fig:p100}).
 
Some aspects of the approach presented in this paper deserve further
studies, and may actually lead to substantial improvements of our 
ability to detect residue contacts from statistical sequence analysis. 
The probably most important question is the capability of our
approach to suppress noise in small MSAs, and to extract contact
information in cases where mean-field DCA fails. This question is
closely related to the determination of an optimal value for the
pattern number $p$ using sequence information alone.
Second, the non-independence of sequences in the alignment,
{\em e.g.} due to phylogenetic correlations, should be taken into 
account in a more accurate way than done currently by sequence 
reweighting. Third, the precise role of the -- heuristically 
determined -- large pseudo-count used to
calculate the Pearson correlation matrix should also be elucidated. 
Fourth, while the use of the Frobenius norm for the coupling 
$e_{ij}(a,b)$ (with the average-product correction, see Methods) has 
proven to be an efficient criterion for contact prediction, it remains
unclear if there exist other contact estimators with better
performance. In this context it would also be interesting to find
a threshold for these contact scores, which separates a signal-rich
from a noise-dominated region. And last but not least, it would be 
interesting to integrate prior
knowledge about proteins, like {\it e.g.}~amino-acid properties or
predicted secondary structure, into the purely statistical inference
approach presented here.

The MATLAB program necessary for the analysis of the data, the
computation of the patterns, and the contact prediction is available
as part of the Supporting Information. Users of the program are kindly
requested to cite the present work.

\section*{Methods}

\subsection*{Data preprocessing} 

Following the discussion of \cite{morcos11}, we introduce two
modifications into the definition Eq.~(\ref{eq:fij}) of the frequency
counts $f_i(a)$ and $f_{ij}(a,b)$:
\begin{itemize}
\item {\it Pseudocount regularization}: Some amino-acid combinations
  $(a,b)$ do not exist in column pairs $(i,j)$, even if $a$ is found
  in $i$, and $b$ in $j$. This would formally lead to infinitely large
  coupling constants, and the covariance matrix $C$ becomes non
  invertible. This divergence can be avoided by introducing a pseudocount 
  $\tilde \nu$, which adds to the occurrence counts of each amino acid
  in each column of the MSA.
\item {\it Reweighting}: The sampling of biological sequences is far
  from being identically and independently distributed (i.i.d.) , it is 
  biased by the phylogenetic history of the
  proteins and by the human selection of sequenced species. This bias 
  will introduce global correlations. To reduce this effect, we decrease 
  the statistical weight of sequences having many similar ones in the 
  MSA. More precisely, the weight of each sequence is defined as the 
  inverse number of sequences within Hamming distance $d_H<xL$, with 
  an arbitrary but fixed $x\in (0,1)$:
  \begin{equation}
    \label{eq:reweighting}
    w_m = \frac 1{||\{ n | 1\leq n\leq M; 
    d_H[(a_1^n,...,a_L^n),(a_1^m,...,a_L^m)]\leq xL \}||} \ 
  \end{equation}
  for all $m=1,...,M$. The weight equals one for isolated sequences, and 
  becomes smaller the denser the sampling around a sequence is. Note that 
  $x=0$ would account to removing double counts from the MSA. The total
  weight
  \begin{equation}
    \label{eq:Meff}
    M_{eff} = \sum_{m=1}^M w_m
  \end{equation}
  can be interpreted as the effective number of independent sequences.
\end{itemize}
With these two modifications, frequency counts become
\begin{eqnarray}
f_i(a) &=& \frac 1{M_{eff}+\tilde\nu}\left[ \frac {\tilde\nu} q +
\sum_{m=1}^M  w_m \; \delta_{a,a_i^m} \right]\\
f_{ij}(a,b) &=& \frac 1{M_{eff}+\tilde\nu} \left[ \frac {\tilde\nu} {q^2} +
\sum_{m=1}^M w_m \delta_{a,a_i^m}\delta_{b,a_j^m} \right] \ .
\label{eq:fij2}
\end{eqnarray}
Values $\tilde \nu \simeq M_{eff}$ and $x\simeq 0.2$ were found to
work optimally across many protein families \cite{morcos11}, we use
these values. Besides these modifications, the Hopfield-Potts-model
learning is performed as explained before.

\subsection*{The number of independent model parameters}

Amino-acid frequencies are not
independent numbers. For instance, on each site $i$, the $q$
amino-acid frequencies add up to one,
\begin{equation}
\sum_{a=1}^{q} f_i(a) =1 \ ,
\label{eq:sumprob}
\end{equation}
and two-site distributions have single-site distributions as
marginals,
\begin{equation}
\sum_{a=1}^{q} f_{ij}(a,b) = f_j(b) \ .
\end{equation}
As a consequence, not all of the constraints (\ref{eq:coherent}) 
and (\ref{eq:f2}) are independent, and the Potts model as given in 
Eq.~(\ref{eq:potts}) has more free parameters than needed to fulfill
the constraints. Families of distinct parameter values result in the 
same model $P(a_1,...,a_L)$ (in physics language, this corresponds 
to a gauge invariance: any function $g_i(a)$ can be added to 
$e_{ij}(a,b)$ and, simultaneously, be subtracted from $h_i(a)$, 
without changing the values of $P$). As in \cite{morcos11}, we 
remove this freedom by setting
\begin{equation}
e_{ij}(a,q) = e_{ij}(q,a) = h_i(q) = 0
\label{eq:gauge}
\end{equation}
for all positions $i,j$ and all amino acids $a$. Within this 
setting, each choice for the parameter values corresponds to a 
different outcome for $P(a_1,...,a_L)$. The parameters
to be computed are therefore the couplings $e_{ij}(a,b)$ and the
fields $h_i(a)$ with $1\le a,b \le q-1$ only. 

An different choice for the gauge is proposed in Text S1, Section 3, and leads to quantitatively equivalent predictions 
for the pattern structures and the contact map.

\subsection*{Mean-field theory for determining the Hopfield-Potts patterns}

The MaxEnt approach underlying DCA can be rephrased in a Bayesian
framework. Assume the model to be given by Eq.~(\ref{eq:potts}), and
assume the sequences in the MSA to be independently and identically
sampled from $P$. The probability of the alignment for given model
parameters (couplings and fields) is then given by
\begin{equation}
P\left[A|\{e_{ij}(a,b),h_i(a)\}\right] = \prod_{m=1}^M P(a_1^m,...,a_L^m)
\ .
\end{equation}
Plugging in Eq.~(\ref{eq:potts}) and defining the log-likelihood of
the model parameters given the MSA $A$, we find
\begin{eqnarray}
{\cal L}\left[ \{e_{ij}(a,b),h_i(a)\} | A \right] &=&
\frac 1M \log P\left[A|\{e_{ij}(a,b),h_i(a)\}\right] 
\nonumber\\
&=& \frac 12\sum_{i,j} \sum_{a,b} e_{ij}(a,b) f_{ij}(a,b) + 
\sum_{i,a} h_i(a)f_i(a) - \log {\cal Z}(\{e_{ij}(a,b),h_i(a)\})
\label{eq:loglikelihood}
\end{eqnarray}
One can readily see that the parameters $\{e_{ij}(a,b),h_i(a)\}$
maximizing ${\cal L}$ are solutions of Eqs.~(\ref{eq:coherent}) and
(\ref{eq:f2}). The corresponding value for the maximum of $\cal L$
coincides with the opposite of the entropy, $-H[P]$, for the MaxEnt
distribution given by Eq.~(\ref{eq:potts}).

Following the study of the Ising model case ($q=2$) in \cite{CoMoSe},
mean-field theory can be used to derive an approximate expression for
the log-likelihood ${\cal L}$ (\ref{eq:loglikelihood}) when the
couplings are chosen to obey Hopfield's prescription,
Eq.~(\ref{eq:hopfield}). Calculations are presented in 
Text S1, Section 1. After optimization over the fields, we are left
with the log-likelihood for the patterns only,
\begin{eqnarray}\label{eq:ll}
&&{\cal L}[\{\boldsymbol\xi\} | A] = {\cal L}_0+\frac 1{2L} 
\sum_{ij,ab} C_{ij}(a,b)\big( \sum_{\mu \le p_+} \xi_{i}^{+,\mu}(a) \xi_{j}^{+,\mu}(b)
- \sum_{\nu \le p_-}\xi_{i}^{-,\nu} (a)\xi_{j}^{-,\nu}(b)\big)
 \\
&+&\frac 12\sum _{\mu \le p_+}\log \left[ 1-\frac 1L \sum _{i,ab} 
\xi_{i}^{+,\mu}(a) C_{ii}(a,b)\xi_{i}^{+,\mu}(b) \right]
+\frac 12\sum _{\nu \le p_-}\log \left[ 1+\frac 1L \sum _{i,ab} 
\xi_{i}^{-,\nu} (a) C_{ii}(a,b)\xi_{i}^{-,\nu}(b) \right]\nonumber
\end{eqnarray}
where ${\cal L}_0=\sum_{i}\sum_{a=1}^q f_i(a) \log f_i(a)$. So we find
the trivial result that, for $p=0$ (no couplings), the log-likelihood 
is the negative of the sum of all single-column entropies, ${\cal L}_0$.  
The optimal patterns, {\em i.e.}~ those optimizing the log-likelihood 
${\cal L}$ are given by Eqs.~(\ref{eq:pattern+}) and (\ref{eq:pattern-}).  
The total log-likelihood corresponding to this selection reads:
\begin{equation}
{\cal L}(p) ={\cal L}_0+  
\sum_{\{\mu| \lambda_\mu \notin [\ell_- , \ell_+ ]\}}
\!\!\!\!  \Delta {\cal L}(\lambda_\mu)\ ,
\end{equation}
where function $\Delta {\cal L}$ is defined in Eq.~(\ref{deltas}), 
and the bounds $\ell_-,\ell_+$ are defined in the Results Section.

The solution given in Eqs.~(\ref{eq:pattern+}) and (\ref{eq:pattern-}) 
is defined up to a rotation in the pattern space, {\em i.e.} up to 
multiplication of all patterns with an indefinite orthogonal 
$(p\times p)$--matrix, ${\cal O}$, in $O(p_+,p_-)$. Indeed,
the patterns $\xi_{i}(a)$ and their rotated counterparts $\hat
\xi _{i}(a)= ( {\cal O} \cdot \xi )_{i}(a)$ define the
same set of couplings $e_{ij}(a,b)$ through
Eq.~(\ref{eq:hopfield}). Note that this invariance is specific
to the Hopfield model, and should not be mistaken for the gauge
invariance of the Potts model discussed in the Results Sections. We
eliminate this arbitrariness according to the following procedure,
detailed in Text S1: Our selection corresponds to
the case where patterns are added one after the other, starting with
the best possible single pattern, followed by the second best
(orthogonal to the first one when single-site correlations
$C_{ii}(a,b)$ are factored out) etc.

\subsection*{Contact prediction from couplings}

Intuitively, residue position pairs with strong direct couplings are
our best predictions for native contacts in the protein structure. To
measure 'coupling strength', we need, however, to map the inferred
$q\times q$ coupling matrices $e_{ij}$ onto a scalar parameter, for each 
$1\leq i<j\leq L$. Whereas previous works on DCA have mainly used the
so-called direct information \cite{weigt09,morcos11}, it was recently
observed that a different score actually improves the contact
prediction starting from the same model parameters $\{e_{ij}(a,b)\}$
\cite{ekeberg12}. To this end, we introduce the Frobenius norm
\begin{equation}
\label{eq:frob_norm}
F_{ij}=\| e'_{ij}\|_2=\sqrt{\sum\limits^{q}_{a,b=1} 
\tilde e_{ij}(a,b)^2}
\end{equation}
of the linearly transformed coupling matrices
\begin{equation}
\label{eq:gauge_change}
\tilde e_{ij}(a,b)= e_{ij}(a,b)-e_{ij}(\cdot,b)-e_{ij}(a,\cdot)+e_{ij}(\cdot,\cdot)\ ,
\end{equation}
where `$\cdot$' denotes average over all amino acids and the gap in
the concerned position. According to the above discussion, this
corresponds to another gauge of the Hopfield-Potts model, more
precisely to the gauge minimizing the Frobenius norm of each coupling
matrix \cite{weigt09}. Further more, the norm is adjusted by an
\textit{average product correction} (APC) term, introduced in
\cite{dunn08} to suppress effects from phylogenetic bias and
insufficient sampling. Incorporating also this correction, we get our
final scalar score:
\begin{equation}
\label{corr_norm}
F^{APC}_{ij}=F_{ij}-\frac{F_{\cdot j}F_{i \cdot}}{F_{\cdot \cdot}}\ ,
\end{equation}
where the '$\cdot$' now indicates a position average.

Sorting column pairs $(i,j)$ by decreasing values of $F^{APC}$
calculated using standard mean-field DCA was shown to give accurate
predictions for residue contacts in various proteins, {\it i.e.}~in
the case where all possible patterns are included ($p= L(q-1)$) in
Eq.~(\ref{eq:hopfield}). The Results Section shows how the performance
in contact prediction varies when the number of patterns is $p\ll
L(q-1)$. 

Note that this criterion gives a coupling score to each pair
of residue positions. The method itself does not provide a cutoff
value for this score, below which predictions should not considered 
any more. Results are therefore typically provided as parametric plots
depending on the number of predicted contacts as a free parameter.

\section*{Acknowledgments}

We are grateful to R. Ranganathan and O. Rivoire for
discussions. 

\newpage
\bibliography{references}

\newpage

\section*{Tables}

{\bf Table 1. Effect of a pattern with two non-zero and opposite components $\xi_i(a)=-\xi_j(b)$.}
\vskip .4cm
\begin{tabular}{c | c | c || c | c } 
$a_i = a?$ &$ a_j = b?$ & $\frac{S(a_1,\ldots , a_L | \boldsymbol\xi )}{\xi_i(a)^2}$ 
& \text{Favored by attractive pattern?} & \text{Favored by repulsive pattern?} \\
 \hline
\text{NO} & \text{NO} & \text{0} & \text{NO} & \text{YES} \\
\text{YES} & \text{NO} & \text{1}& \text{YES} & \text{NO}  \\
\text{NO} & \text{YES} & \text{1} & \text{YES} & \text{NO} \\
\text{YES} & \text{YES} & \text{0} & \text{NO} & \text{YES} \\
\end{tabular}

\vskip 1cm
{\bf Table 2. Effect of a pattern with two non-zero and equal components $\xi_i(a)=\xi_j(b)$.}
\vskip .4cm
\begin{tabular}{c | c | c || c | c } 
$a_i = a ?$& $a_j = b ?$&$ \frac{S(a_1,\ldots , a_L | \boldsymbol\xi )}{\xi_i(a)^2}$ 
& \text{Favored by attractive pattern?} & \text{Favored by repulsive pattern?} 
\\ \hline
\text{NO} & \text{NO} & \text{0} & \text{NO} & \text{YES} \\
\text{YES} & \text{NO} & \text{1}& \text{NO} & \text{YES} \\
\text{NO} & \text{YES} & \text{1} & \text{NO} & \text{YES} \\
\text{YES} & \text{YES} & \text{4} & \text{YES} & \text{NO} \\
\end{tabular}

\ \vspace{1cm}

\section*{Supporting Information Files}

{\bf Text S1.}  Supporting Information for From principal component to direct coupling analysis
of coevolution in proteins: Low--eigenvalue modes are needed for
structure prediction.\\

\noindent {\bf Code S1.} Matlab code for the Hopfield-Potts inference.

\newpage
\section*{Figures and Legends}

\newpage

\end{document}